\documentclass[%
 aip,
 amsmath,amssymb,
 reprint,%
]{revtex4-1}
\usepackage {CJK}

\newcommand*{\citen}[1]{%
  \begingroup
    \romannumeral-`\x 
    \setcitestyle{numbers}%
    \cite{#1}%
  \endgroup   
}

\usepackage{graphicx}
\usepackage{dcolumn}
\usepackage{bm}

\usepackage[utf8]{inputenc}
\usepackage[T1]{fontenc}
\usepackage{mathptmx}

\usepackage{subfig}
\graphicspath{{figures/}}

\newcommand{\norm}[1]{\| #1 \|}
\newcommand{\ud}{\mathrm{d}}
\newcommand{\oh}{\frac{1}{2}}
\newcommand{\mb}[1]{\mathbf{#1}}

\usepackage{hyperref}
\usepackage{soul} 
\usepackage{xcolor} 
\setstcolor{red}
\newcommand{\edit}[1]{\textcolor{black}{#1}}

\usepackage{caption}

\begin{document}
\preprint{AIP/123-QED}

 \begin {CJK*} {GB} { }
 
\title[Leaky Cell Model]{Leaky Cell Model of Hard Spheres}

\author{Thomas G.~Fai}
 \email{tfai@brandeis.edu}
\affiliation{ 
\mbox{Department of Mathematics and Volen Center for Complex Systems, Brandeis University, Waltham, MA, USA} 
}%

\author{Jamie M.~Taylor}
\affiliation{%
Basque Center for Applied Mathematics (BCAM), Bilbao, Bizkaia, Spain
}%

\author{Epifanio G.~Virga}
\affiliation{%
Dipartimento di Matematica, Universit\`{a} di Pavia, Pavia, Italy
}%

\author{Xiaoyu Zheng}
\author{Peter Palffy-Muhoray}%
  \altaffiliation[Also at ]{Advanced Materials and Liquid Crystal Institute, Kent State University, Kent, OH, USA}
\affiliation{ 
Department of Mathematical Sciences, Kent State University, Kent, OH, USA
}%

\date{\today}

\begin{abstract}
We study packings of hard spheres on lattices. The partition function, and therefore the pressure, may be written solely in terms of the accessible free volume, i.e.~the volume of space that a sphere can explore without touching another sphere. We compute these free volumes using a leaky cell model, in which the accessible space accounts for the possibility that spheres may escape from the local cage of lattice neighbors. We describe how elementary geometry may be used to calculate the free volume exactly for this leaky cell model in two- and three-dimensional lattice packings and compare the results to the well-known Carnahan-Starling and Percus-Yevick liquid models. We provide formulas for the free volumes of various lattices and use the common tangent construction to identify several phase transitions between them in the leaky cell regime, indicating the possibility of coexistence in crystalline materials.
\end{abstract}

\maketitle
 \end{CJK*}

\section{\label{sec:intro}Introduction}
The thermodynamical properties of hard spheres, i.e.~objects whose interior points cannot overlap with those of any other hard object, have been a subject of great theoretical interest given the conceptual simplicity and rich phenomena exhibited by these systems \cite{torquato2010jammed}. Systems of hard spheres are one of the few systems that are analytically tractable, allowing for exact and/or approximate calculations of physical properties \edit{as reviewed in \citen{lowen2000fun,leefrenkel,santos2020structural}}. Despite the relative simplicity of hard spheres, they exhibit non-trivial behaviors such as phase transitions and allow for experimentally testable predictions\edit{\cite{santos2016concise}}.

This subject has experienced a resurgence because of new computational tools for many-body problems. Computer simulations, first used to generate random packings of hard disks in two dimensions by the method of random sequential addition \cite{hoover1979exact} and to calculate freezing/melting transitions by dynamic Monte Carlo simulation \cite{hoover1967use}, have been used to study packings of circles and ellipses in two dimensions \cite{vieillard1972phase} and spheres, ellipsoids, and other shapes in three dimensions \cite{chaikin2006some,ni2012phase}, in tandem with sophisticated computational methods to analyze packings based on techniques such as computational topology \cite{carlsson2012computational} and discrete geometry \cite{sastry1997statistical,sastry1998free,kapfer2012jammed,maiti2014free,chen2017random}. Boltzmann generators based on deep learning have recently been developed to sample equilibrium states of confined systems of hard disks \cite{noe2019boltzmann}.

This resurgence of interest and new data has highlighted the need for predictive theories grounded in statistical mechanical first principles. A key quantity that appears throughout these works is the \textit{free volume}, i.e.~the volume of space available to the center of mass of each particle in the packing.
The partition function may be approximated in terms of the free volume accessible to each sphere, which only requires a purely local calculation in terms of a cage of neighboring spheres. In the special case of lattice packings with a prescribed crystal structure, only a single unit cell needs to be considered, and this observation led to the development of a cell theory (CT) of hard sphere liquids \cite{eyring1937theory,hirschfelder1937theory,lennard1937critical,lennard1938critical}.

This cell theory (alternatively called the Lennard-Jones-Devonshire cell model) allows for various analytically tractable calculations. One can derive exact expressions for the free volume in the case that the spheres are arranged in a lattice\cite{buehler1951free} and use these formulas in the partition function. In this work, we further explore the cell theory by computing expressions for the free volume for several crystal structures in one, two, and three dimensions. Whereas the cell theory has thus far primarily been used to model solids in the high-density limit, here we introduce a \textit{leaky cell model} which represents materials that retain their crystal structure at low density. We use the resulting expressions to predict phase transitions between lattices and interpret these predictions in the context of numerical and experimental observations.

The extension of cell theories to low densities was already considered in Buehler et al.~\cite{buehler1951free}, in which they average between \emph{hard center} and \emph{soft center} cell models to capture the possibility that spheres may wander out of their local cages at sufficiently low densities. However, in all the scenarios they considered, the center of the wandering sphere is always restricted to a local Wigner-Seitz (or Voronoi) cell. \edit{In \citen{buehler1951free}, spheres are taken to be ``bounded either by the collision spheres of its nearest neighbors or else by mathematical partitions which bisect the distance between lattice points, whichever barrier appears first.'' It is natural then to consider what happens when these mathematical partitions are absent.}

\edit{It has been shown that lattice structures lose mechanical stability at low density \cite{woodcock1997computation,warshavsky2018mechanical}. However, the ``ghost lattice'' obtained at low density is nevertheless a useful theoretical instrument which allows us to predict behaviors such as phase transitions within a fully consistent cell theory. Here, we further develop this concept by pushing the cell theory into the low and intermediate-density regime in which spheres may leak out of their local cages.}

In the leaky cell model proposed here, we allow for the possibility that spheres are able to explore an extended region outside of the Wigner-Seitz cell. We identify an intermediate regime in which spheres escape the Wigner-Seitz cell---i.e. they may be brought outside the cell by a continuous translation without making contact with any other sphere---and we find the critical packing fractions at which the leaky cell model begins to deviate from the classical one. We consider face-centered cubic (FCC), body-centered cubic (BCC), and simple cubic (SC) lattices, and we find phase transitions between these lattices in their corresponding leaky regimes.

To characterize the behavior of lattice packings in this leaky, moderate density regime, we perform exact calculations of the free volume. Given its analytic tractability, the leaky cell model extends lattice models outside their typical range of application and yields predictions for crystalline materials at moderate densities. The formulas for free volume that we compute have interesting features such as discontinuities in \edit{the compressibility factor} observed for certain lattices. By allowing for a direct comparison of the free volume of several lattice packings, this work elucidates the thermodynamical importance of this quantity and provides a foundation for the study of packings with local or global crystal structure.

We further consider an apparent limitation of the cell theory. Whereas using the cell theory to model the high-density (solid) regime and empirical equations of state to model the low-density (liquid) regime leads to good agreement with observations from experiment and Monte Carlo simulation, our results indicate that equations of state based purely on cell theory are unable to accurately predict the freezing/melting transitions in a hard sphere gas. We rectify this by introducing a new quantity, called the \textit{quasi-communal entropy}, which characterizes the difference between the hard sphere gas and cell theory in the low-density regime.

This article is structured as follows. First, we describe the formulation of the leaky cell model and its application to several three-dimensional (3D) lattices. (The two-dimensional (2D) and one-dimensional (1D) cases are also described in the appendices.) Next, we explain how to derive exact formulas for the corresponding free volume measures based on a straightforward geometric argument. Given these formulas, it is straightforward to compute various physical quantities such as the free energy and \edit{compressibility factor}. We use the common tangent construction to identify packing fractions at which different lattices may coexist with one another or with liquids described by the empirical Percus-Yevick and/or Carnahan-Starling equations of state.

\section{\label{sec:leak}Leaky cell model}
We calculate free volumes exactly in regular lattice arrangements such as cubic lattices and hexagonal close-packed (HCP) lattices. The number of spheres and total volume are fixed. The locations of all but one sphere are fixed on the lattice. The free volume is defined to be the volume of the region accessible to the center of mass of a sphere that wanders continuously from its lattice site without making contact with any other sphere \cite{hoover1979exact}.

\subsection{\edit{Compressibility factor} from free volume}
In a hard sphere gas, the equation of state may be determined purely from the free volume, as we now derive from first principles following closely the reasoning of previous works \cite{rice1944statistical,kirkwood1950critique,wood1952note,speedy1991cavities}. Once we have a formula $\mathcal{F}=\mathcal{F}(v)$ for the free volume $\mathcal{F}$ as a function of the Voronoi cell volume $v$, we construct the partition function $\mathcal{Z}$ in terms of $\mathcal{F}$.

For $N$ hard spheres with center of mass positions $\mb{x}_i$ and momenta $\mb{p}_i$ for $i=1,\dots,N$, the partition function in the case of single occupancy of each lattice cell is given by\cite{eyring1937theory}
\begin{align}
  \mathcal{Z} &= \frac{1}{h^{3N}}\idotsint \ud \mathbf{x}_1 \dots \ud \mathbf{x}_N \idotsint \exp{\left( \frac{- \norm{\mathbf{p}_i}^2}{2 m k_B T}\right)} \ud \mathbf{p}_1 \dots \ud \mathbf{p}_N \notag \\
&=\left(2 \pi m k_B T/h^2\right)^{3N/2} \idotsint \ud \mathbf{x}_1 \dots \ud \mathbf{x}_N \notag \\
&=\left(2 \pi m k_B T/h^2\right)^{3N/2} \mathcal{F}^N,
\label{eq:pfun}
\end{align}
where the free volume $\mathcal{F}$ depends on the lattice configuration and the excluded volumes of neighboring spheres, $h$ is Planck's constant, $k_B$ is Boltzmann's constant, $T$ is the temperature, and $m$ is the mass of each sphere. (Note that the usual factor of $1/N!$ does not appear explicitly in the denominator of \eqref{eq:pfun}. This is because there is another factor of $N!$ in the numerator to account for all permutations of spheres among lattice sites, and these two factors exactly cancel.) Therefore
\begin{equation}
\label{eq:part1}
\log\mathcal{Z} = N\log\mathcal{F}-3N\log\Lambda,
\end{equation}
where we have introduced the de Broglie wavelength ${\Lambda = h/\sqrt{2\pi m k_B T}}$. The Helmholtz free energy $A$ is defined in terms of $\mathcal{Z}$ via:
\begin{equation}
A := -k_B T\log\mathcal{Z}.
\end{equation}
Substituting the expression \eqref{eq:part1} for $\mathcal{Z}$ into this equation yields
\begin{equation}
\label{eq:free1}
A = -k_B T\left(N\log\mathcal{F}-3N\log\Lambda\right).
\end{equation}
The pressure $p$ is defined by
\begin{equation}
p := -\frac{\partial A}{\partial V},
\end{equation}
i.e.~the pressure is the negative gradient of the free energy with respect to the total volume (at constant temperature). The second term in \eqref{eq:free1} is independent of volume, so that differentiating yields
\begin{equation}
p= k_B T\frac{\partial \log\mathcal{Z}}{\partial V}=\frac{N k_B T}{\mathcal{F}} \frac{\partial \mathcal{F}}{\partial V}.
\end{equation}
The Voronoi cell volume $v$ and total volume $V$ are related on the lattice by $V=Nv$, so that the pressure may be rewritten in terms of local quantities as
\begin{equation}
p=\frac{k_B T}{{\mathcal{F}}} \frac{\partial \mathcal{F}}{\partial v}.
\end{equation}
Finally, the compressibility factor $Z$ is defined via
\begin{equation}
Z:=\frac{pV}{N k_B T},
\end{equation}
and can therefore be expressed in terms of the free volume by
\begin{equation}
Z=\frac{v}{\mathcal{F}}\frac{\partial \mathcal{F}}{\partial v}.
\label{eq:cf}
\end{equation}
The above formula, which also appears in Buehler et al.~\cite{buehler1951free}, provides the foundation for our calculations. We use it to compute the \edit{compressibility factor} $Z$ directly from the free volume $\mathcal{F}$ and the number density $\rho = v^{-1}$.

\section{Results}
We next describe how the free volumes of lattices of hard spheres may be computed analytically for the leaky cell model. We will denote the packing fraction by $\eta$ throughout the remainder of the article. Letting $v_0$ denote the volume of each sphere in the system, $\eta=v_0/v$.

\subsection{Exact formulas on lattices}
The free volume in cell theory is defined in terms of the region accessible to the center of mass of a sphere holding the position of every other sphere fixed. It is the volume of the region accessible by continuous translation without making contact with another sphere--this is a connected component in the set of points located a distance of at least $2R$ from any other sphere, where $R$ is the sphere radius.

Given that lattice structures are typically associated with solids, the cell model has classically been applied in the high-density limit, in which case a sphere cannot escape from the unit cell without making contact with a neighboring sphere, which is forbidden in the hard sphere model. Therefore, to compute the free volume in the high-density limit it is sufficient to consider the unit cell only. In the \textit{leaky cell model} explored here, we extend the cell theory to lower packing fractions in which spheres are able to escape from the unit cell.
\begin{figure}
\hspace*{-0.4cm}\includegraphics[trim=0 0 .3in 0, clip,width=0.55\textwidth]{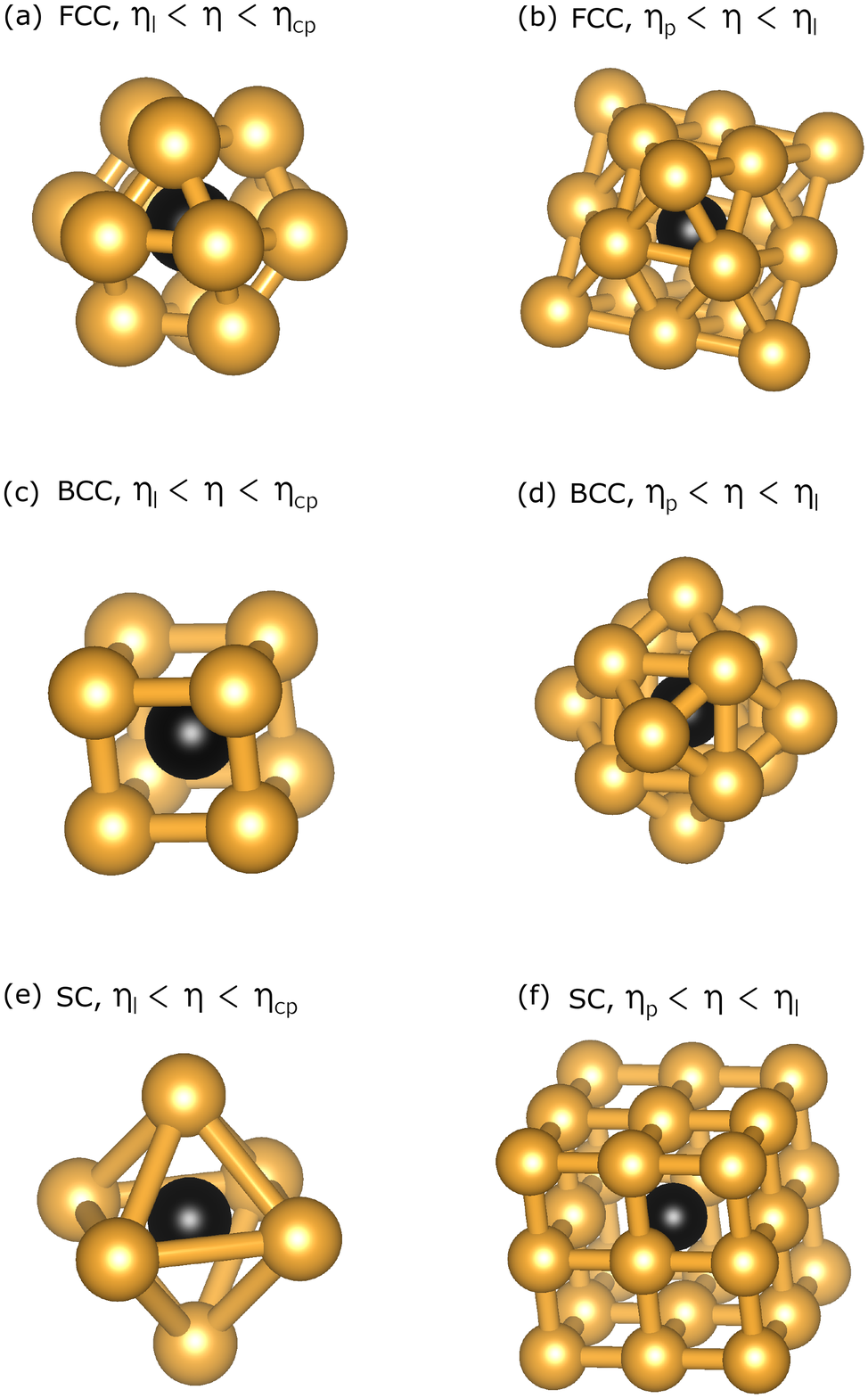}
\caption{\label{fig:loc_pack} Development of the leaky cell model. Each polyhedron represents the cage of neighbors for a chosen sphere in the lattice, with neighboring spheres making up the vertices of the polyhedron. The left column illustrates the cage at high packing densities near the close-packing fraction $\eta_\text{cp}$, whereas the right column illustrates the extended cage of spheres between the leaky packing fraction $\eta_\text{l}$ and the percolation limit $\eta_\text{p}$. Cell structures were visualized using VESTA \cite{momma2011vesta}.}
\end{figure}

\subsubsection{FCC/HCP lattice}
We first describe packings of spheres in the FCC lattice, which involves planar layers of hexagonally-arranged spheres. Note that, as pertains to the free volume, the HCP and FCC lattices are equivalent; although we discuss only the FCC lattice, the resulting formulas apply to the HCP lattice as well. (In the HCP lattice, the layers are stacked in an ABAB pattern, whereas, in the FCC lattice they are stacked in an ABCABC pattern. It follows that the FCC lattice and HCP lattice are equivalent up to rotating one half of the unit cell, so that the resulting free volume formulas are the same.)

As we discuss in detail later in this paper, for packings denser than the leaky packing fraction $\eta^\text{FCC}_\text{l}$, the formulas we derive are equivalent to those given previously by \citen{buehler1951free}. They differ for packings less dense than $\eta^\text{FCC}_\text{l}$ because in the leaky cell model we must also account for the accessible volume outside the Wigner-Seitz cell. 

The void space surrounding a sphere in the lattice can be partitioned into $N_t=8$ tetrahedra and $N_o=6$ octahedra. As the packing fraction increases, spheres becomes caged first by 18 neighboring spheres forming a polyhedron with triangular octahedral faces and subsequently by 12 neighboring spheres forming a \edit{polyhedron} with both triangular and square faces (the octahedral midplanes). See Figure \ref{fig:loc_pack}(a)--(b).
\begin{figure*}
\includegraphics[width=\textwidth,trim=0in 4.3in 0in 0in, clip]{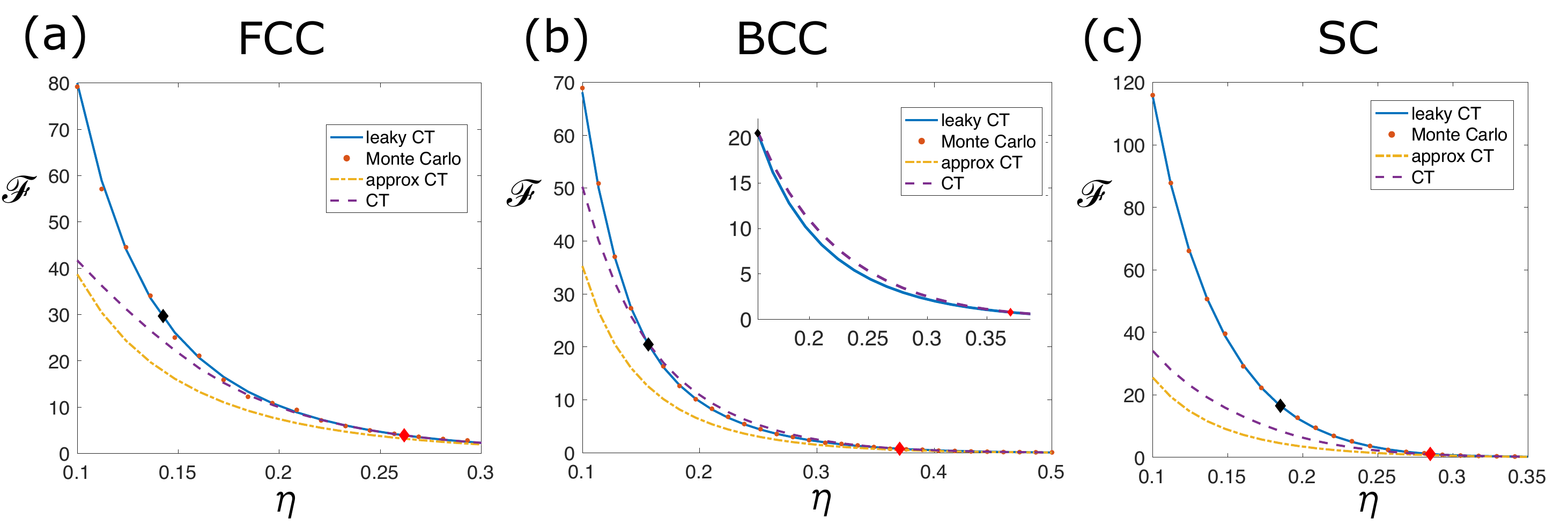}
\caption{\label{fig:comp}Comparison between the free volumes in the leaky cell model and the classical cell theory\cite{buehler1951free}. The asymptotic approximation corresponding to the curves labeled ``approx CT'' is given in Appendix \ref{app:approx}, the black and red diamonds indicate the percolation and leaky packing fractions, respectively, of Table~\ref{tab:fractions}, and the points labeled Monte Carlo are obtained by sampling points uniformly within the cage and rejecting any that are inside of a neighboring exclusion sphere. (a) FCC/HCP lattice, (b) BCC lattice, (inset) Zoom-in showing that the free volume of the leaky cell model is bounded above by the classical cell theory between the percolation and leaky cell transitions, and (c) SC lattice.}
\end{figure*}
As mentioned above, there is a transition in the free volume as the spheres lose the ability to escape through the square faces at octahedral midplanes. This leaky cell transition occurs when the edge length $a$ satisfies $a=2\sqrt{2} R$, which corresponds to the packing fraction $\eta^\text{FCC}_\text{l} \approx 0.26$. Interestingly, this is a smooth transition; there is no discontinuity in the free volume or its first derivative.

The percolation transition $\eta^\text{FCC}_\text{p}$ takes place at a smaller packing fraction at which the spheres cannot escape through triangular faces. Assuming the spheres have radius $R$, this occurs when $a=2\sqrt{3} R$, and at this point spheres become locally caged and unable to move freely throughout the domain. The packing fraction at which this occurs is $\eta^\text{FCC}_\text{p} = \Omega/(3\sqrt{3/2})\approx 0.15$, where $\Omega=\cos^{-1}(23/27)$ is the solid angle subtended by each vertex of a regular tetrahedron.

The free volume is defined as a piecewise function between these transitions, with the formula given in Eq.~\eqref{eq:fv_hcp} of Appendix \ref{app:exact}. As detailed there, the formula we use involves the three-dimensional double intersection volume $V_{2I}(a)$ of intersecting spheres, as well as the intersection volume of three and four spheres.
In order to calculate these intersection volumes, we use general formulas for volumes of higher-order intersections as given in \citen{gibson1987exact,gibson1987volume}.

In Fig.~\ref{fig:comp}(a) we show that at high density our results agree with those of \citen{buehler1951free}. The disagreement at low densities is because of different formulations are used for the free volume: in \citen{buehler1951free} the free volume is taken to be a subset of the dodecahedral Voronoi cell associated to a given sphere, whereas in our case we include all accessible volume in the lattice including the voids that emerge when $a \ge 2\sqrt{2} R$ so that spheres can escape through square faces of the cuboctahedron associated with the FCC lattice but remain caged by triangular faces.

\subsubsection{BCC lattice}
There are notable differences in the leaky cell model for the body-centered cubic (BCC) lattice. Initially, spheres are caged by 8 spheres at the vertices of a cube (Figure \ref{fig:loc_pack}(c)--(d)). At the leaky cell fraction $\eta^\text{BCC}_\text{l} \approx 0.37$, spheres become caged by 14 spheres, which include the 6 spheres located across cubic faces in addition to the 8 spheres at cubic vertices. Below the leaky cell fraction, the void space may be tesselated by 6 octahedra. The free volume is computed by subtracting the exclusion spheres of the 14 neighbors from the volumes of these octahedra, accounting for all intersections (going up to quintuple intersections in this case). See Appendix \ref{app:exact} for the explicit free volume formulas.

Unlike the FCC and SC lattices, in which spheres can escape from their local cages at the leaky packing fraction, spheres in a BCC lattice are \textit{unable} to escape from their cubic cage at the leaky cell fraction. This is because the neighboring spheres in the lattice positioned on the opposite sides of the faces of the cubic cage also exclude volume within the cube. Consequently, the free volume for the leaky cell model is actually less than the free volume predicted by the cell theory (Fig.~\ref{fig:comp}(b)).
\begin{figure*}
\includegraphics[width=\textwidth,trim=0in 4.5in 0in 0in, clip]{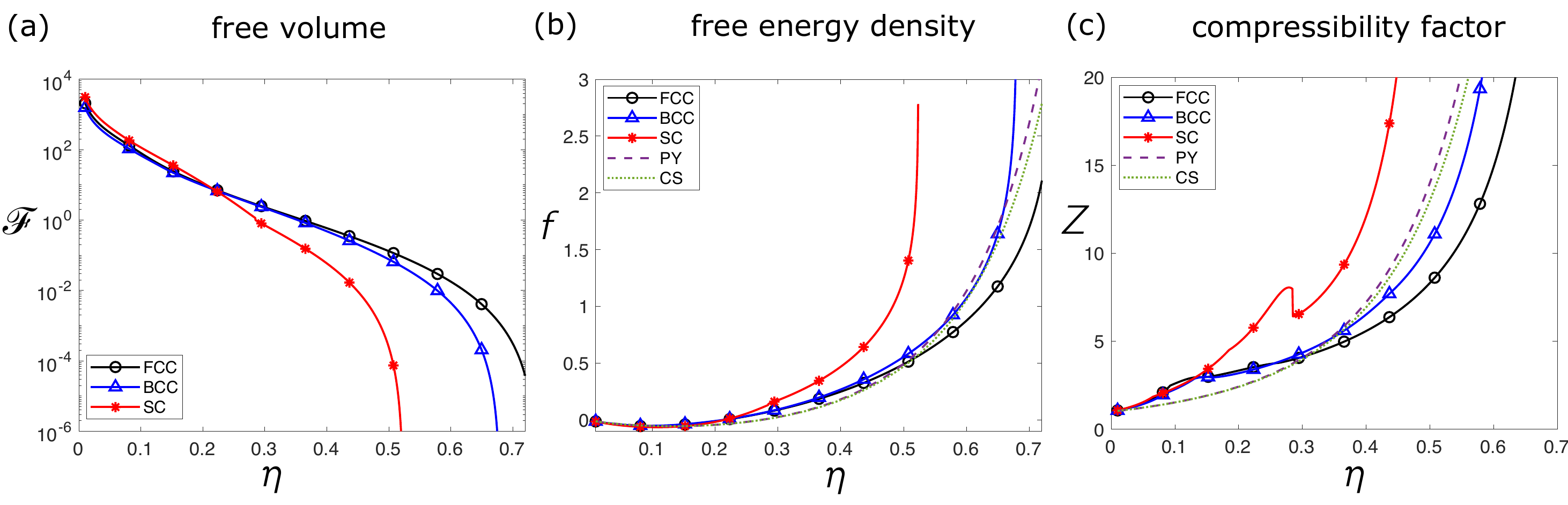}
\caption{\label{fig:cf} Properties of the leaky cell model for FCC/HCP, BCC, and SC lattices. (a) Free volume $\mathcal{F}$ in units of $R^3$ as a function of the packing density $\eta$, (b) free energy density $f$ vs.~the packing density $\eta$, (b) \edit{compressibility factor} $Z$ vs.~the packing density $\eta$.}
\end{figure*}

\subsubsection{SC lattice}
We next consider the simple cubic lattice. In addition to the percolation threshold at $\eta^\text{SC}_\text{p} \approx 0.19$, at which diagonal neighbors on square faces begin to intersect and the center sphere becomes caged by 26 neighbors, the SC lattice has a transition at $\eta^\text{SC}_\text{l} \approx 0.29$ beyond which the sphere can no longer escape through triangular faces (Figure \ref{fig:loc_pack}(e)--(f)), so that opposite corners of cubes are lost and the number of neighbors decreases to 6.

\edit{In Appendix \ref{app:2D}, we use analogous geometrical arguments on lattices in 2D to compute free volumes and the resulting compressibility factors.}

\subsection{Comparison between different lattices}
The critical packing fractions of the 3D lattice models are summarized in Table~\ref{tab:fractions}.
\begin{table}
\caption{\label{tab:fractions}Percolation ($\eta_\text{p}$), leaky ($\eta_\text{l}$), and close-packed ($\eta_\text{cp}$) packing fractions in the FCC, BCC, and SC lattices.}
\begin{ruledtabular}
\begin{tabular}{lccr}
3D Lattice&$\eta_\text{p}$&$\eta_\text{l}$&$\eta_\text{cp}$\\[0.4em] \hline\\[-0.8em]
FCC&$\frac{\pi}{9\sqrt{6}}\approx 0.14$&$\frac{\pi}{12}\approx 0.26$&$\frac{\pi}{3\sqrt{2}}\approx 0.74$ \\[0.8em]
BCC&$\frac{9\pi}{128\sqrt{2}}\approx 0.16$&$\frac{\pi}{6\sqrt{2}}\approx 0.37$&$\frac{\pi\sqrt{3}}{8}\approx 0.68$\\[0.8em]
SC&$\frac{\pi}{12\sqrt{2}}\approx 0.19$&$\frac{2\pi}{9\sqrt{6}}\approx 0.29$&$\frac{\pi}{6}\approx 0.52$
\end{tabular}
\end{ruledtabular}
\end{table}

As mentioned previously, above the leaky packing fraction the leaky cell model reduces to the classical cell theory. To illustrate the difference between the models below the leaky packing fraction, in Fig.~\ref{fig:comp} we compare the free volumes (in units of $v_0$) obtained by either the leaky cell model or by classical cell theory. Interestingly, the differences between the leaky and classical models depend strongly on the lattice. In particular, there are qualitiative differences between the FCC and BCC lattices; in the FCC lattice, the leaky cell model has a free volume that is bounded below by the classical cell theory, whereas in the BCC lattice it is bounded above by the classical cell theory for $\eta>\eta^\text{BCC}_\text{p}$. This is a direct consequence of the lattice geometry. In the FCC lattice, the wandering sphere leaks out of the unit cell, whereas in the BCC lattice the exclusion spheres of neighbors outside the unit cell leaks in.

In Fig.~\ref{fig:cf}, we plot the free volumes (in units of $R^3$), free energy densities (in $k_B T$ units), and \edit{compressibility factors} obtained for the three lattice models described above. Fig.~\ref{fig:cf}(c) includes a comparison of the \edit{compressibility factor} $Z$ from these lattice free volumes---computed according to Eq.~\eqref{eq:cf}---to standard liquid models such as Percus-Yevick and Carnahan-Starling for which
\begin{align}
Z^c_{PY}(\eta) &= \frac{1+\eta+\eta^2}{(1-\eta)^3} \\
Z_{CS}(\eta) &= \frac{1+\eta+\eta^2-\eta^3}{(1-\eta)^3},
\end{align}
where $Z^c_{PY}$ and $Z_{CS}$ denote the Percus-Yevick and Carnahan-Starling \edit{compressibility factors}, respectively. \edit{Note that the Percus-Yevick approximation does not lead to a unique compressibility factor; the superscript in $Z^c_{PY}$ denotes that this compressibility factor is obtained through the compressibility route (as opposed to the energy, virial, or chemical-potential routes \cite{santos2016concise}). For convenience, in what follows we will refer to this expression obtained through the compressibility route simply as the Percus-Yevick approximation.} Note the jump discontinuity in $Z$ for the simple cubic lattice that arises because $\mathcal{F}(v)$ is not continuously differentiable at $\eta^\text{SC}_\text{l}$.

\begin{figure*}
\hspace*{-.4in}\includegraphics[width=1.1\textwidth,trim=0in 1.7in 0in 0in, clip]{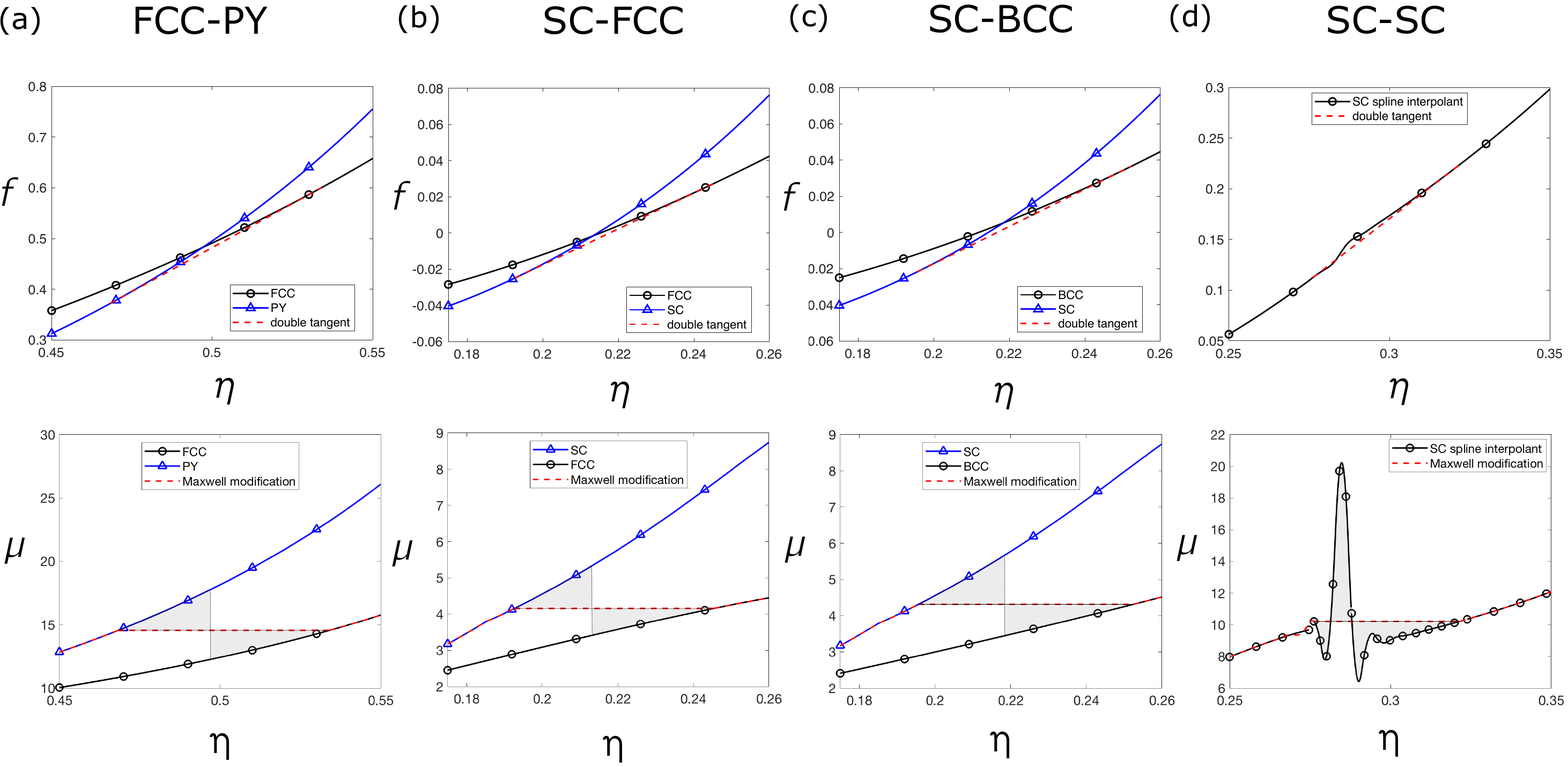}
\caption{\label{fig:freez}Phase transitions in the leaky cell model. Top row: common tangent construction, bottom row: Maxwell construction via the chemical potential $\mu$, in which the shaded areas represent equal areas. (a) We recover the known \edit{PY-FCC} freezing/melting transition. There is an analogous \edit{PY-BCC} freezing transition at $\eta \approx 0.55$ (not shown). (b) SC-FCC phase transition at $\eta \approx 0.22$, (c) SC-BCC phase transition at $\eta \approx 0.22$, (d) SC-SC phase transition at $\eta \approx 0.3$. A polynomial interpolant is used to smooth out the discontinuity in the free energy density that occurs at $\eta_\text{l}^\text{SC}$.}
\end{figure*}
\subsection{Coexistence}
We apply the common tangent construction to the free energy densities obtained by the leaky cell model in order to determine the existence of phase transitions. We first \edit{recover the theoretical freezing/melting transition} between the liquid phase and FCC lattice at packing fractions of approximately 0.47 \& 0.53 respectively (with 0.47 corresponding to the liquid packing fraction and 0.53 corresponding to the solid packing fraction) \cite{hoover1968melting}. \edit{Note that the empirical values of the freezing and melting packing fractions are actually closer to 0.49 and 0.54 \cite{hoover1968melting}.} As shown in Fig.~\ref{fig:freez}(a) and Table~\ref{tab:freez},
\begin{table}
\caption{\label{tab:freez}Phase transitions in the leaky cell model.}
\begin{ruledtabular}
\begin{tabular}{ll}
Phase transition&Corresponding packing fractions
\\[0.4em] \hline\\[-0.8em]
PY--FCC&0.47--0.53\\
PY--BCC&0.54--0.57\\
SC--FCC&0.19--0.25\\
SC--BCC&0.20--0.25\\
SC--SC (quasi)\footnote{We regard this as a quasi-phase transition, since only a near common tangent is observed.}&0.28--0.32\\
PY--SC&none\\
BCC--FCC&none
\end{tabular}
\end{ruledtabular}
\end{table}
we are able to recover this result. Since this phase transition occurs above the leaky packing fraction of $\eta^\text{FCC}_\text{l} \approx 0.26$, the leaky cell model simply recovers the freezing/melting transition predicted by the classical cell theory. In addition to the common tangent construction, we also verify our result using the equivalent equal area Maxwell construction \cite{binder2012beyond}. See Fig.~\ref{fig:freez}. For the BCC lattice, we find a \edit{PY-BCC} freezing/melting transition at packing fractions of 0.54 \& 0.57 analogous to the known \edit{PY-FCC} transition. As discussed next, using the appropriate calibration to high-density is critical for obtaining this phase transition.

We find SC--FCC and SC--BCC inter-lattice phase transitions that occur at packing fractions of approximately 0.19 \& 0.24, respectively, which is between the leaky packing fraction and the percolation limit (Figs.~\ref{fig:freez}(b) and \ref{fig:freez}(c)). This indicates that different lattice structures may coexist in the leaky regime. Further, we find an SC--SC quasi intra-lattice phase transition around packing fractions of 0.29 \& 0.33, which is again within the leaky regime. Upon close inspection of the free energy density for the SC lattice, one finds that there is a near common tangent between these packing fractions, but that there is only near-tangency, i.e.~the tangent misses the curve by approximately 0.5\%. We have verified this phase transition by constructing the convex hull of the free energy, an approach that generalizes to non-differentiable functions such as the present one.

Moreover, we can rule out the existence of phase transitions between other lattice structures and between Percus-Yevick or Carnahan-Starling liquids and the SC lattice. This is because the associated free energy densities do not intersect, i.e.~they are bounded above or below by one another so there can be no common tangent (Fig.~\ref{fig:cf}(b)).

\begin{figure*}
\includegraphics[width=\textwidth,trim=0in 1.2in 0in 0in, clip]{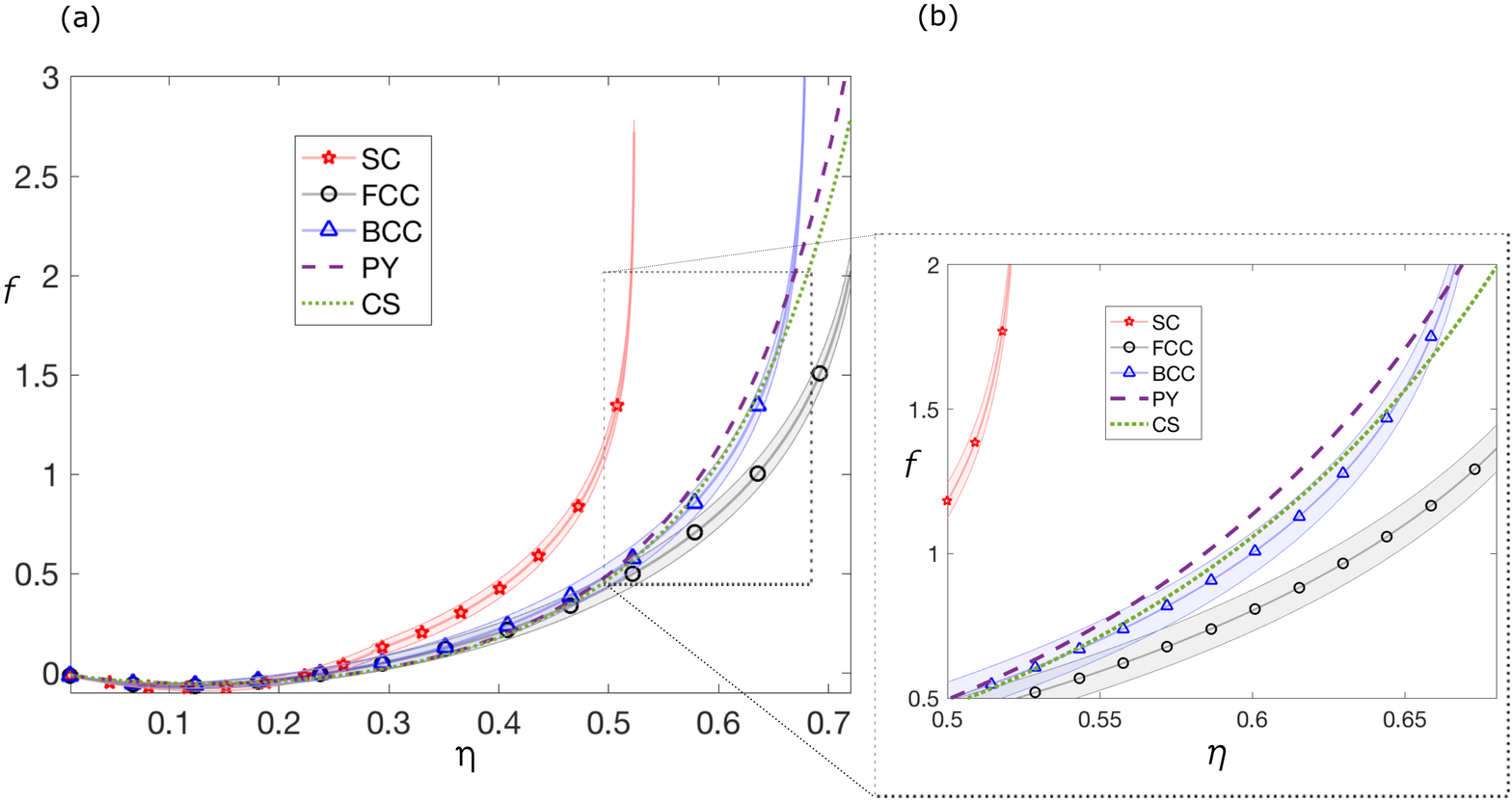}
\caption{\label{fig:range} (a) There is an ambiguity in the absolute scale of the free energy densities of the leaky or classical cell theories depending on the model assumptions \cite{eyring1937theory}. However, with the exception of the BCC model (see inset (b)) the existence of a phase transition does not depend on whether the high-density or low-density calibration is used.}
\end{figure*}
\subsection{Calibration}
The free energy density is calculated in terms of the exact free volume through Eq.~\eqref{eq:free1}, which invokes the single occupancy assumption and therefore can only expected to be accurate in the high-density regime. If one replaces the single-occupancy assumption by allowing movement from cell to cell and positing that the total free volume is simultaneously accessible to all spheres as it would be in a low-density gas, the partition function is increased by a factor of $e$ per molecule \cite{eyring1937theory}. The resulting expressions for the free energy density differ only up to an additive constant, and because this additive constant is the same across all lattice models (provided they are treated consistently) it does not affect the predicted phase transitions. However, the choice of additive constant \textit{does} affect the phase transitions when lattice models are compared to other free energy densities on an absolute scale, e.g.~the Percus-Yevick or Carnahan-Starling liquids. From a mathematical standpoint, this is because the common tangent construction that defines these phase transitions involves the absolute free energy density and not only its derivatives.

More generally, the free energy is often obtained by integrating the pressure from a reference value \cite{ponce1976analytical,mirzaeinia2017equation}:
\begin{align}
\label{eq:free_en_int}
\beta \frac{A}{N}-\beta \frac{A^\text{IG}}{N} = \int_0^\eta \frac{Z-1}{\eta} \ud \eta,
\end{align}
in which $A^\text{IG}$, the free energy of an ideal gas, satisfies $A^\text{IG} = -1+3\log{\frac{\Lambda}{2R}}+\log{\frac{6\eta}{\pi}}$. These two approaches for normalizing the free energy, i.e.~whether one directly uses the absolute free energy given by \eqref{eq:free1} or integrates the \edit{compressibility factor} from a low-density reference according to \eqref{eq:free_en_int}, are essentially a question of whether one calibrates the free energy using either the low-density or high-density reference value, an issue also considered by previous authors \cite{alder1968studies}.

We find that the phase transitions may depend sensitively on this calibration, and in some cases this calibration affects whether or not a transition exists at all. Whereas the \edit{PY-FCC} phase transition does not depend on this particular assumption---i.e.~regardless of whether the free energy resulting from Eq.~\eqref{eq:free1} is used or whether a factor of $e$ is added to the free energy density, the same phase transitions exist---the actual location of the phase transition varies by as much as 10\%. Moreover, the result is robust to the choice of the empirical model for the liquid state (Fig.~\ref{fig:range}).

In contrast, the freezing/melting transition in the BCC lattice is highly sensitive to the choice of empirical model. If the free energy is calibrated to high-density and the Carnahan-Starling equation of state is used, no freezing/melting transition is predicted, whereas using the low-density calibration or the Percus-Yevick liquid model leads to the existence of a freezing transition at packing fractions of approximately 0.55--0.65 (Fig.~\ref{fig:range}(b)).

Figure \ref{fig:range} illustrates the range of free energy densities obtained through the leaky cell model depending on whether one includes the additive constant. It shows that this ambiguity does not affect whether a freezing/melting transition exists between the FCC lattice and the Percus-Yevick or Carnahan-Starling liquids. Further, it does not affect any intra- or inter-lattice phase transitions, since the additive constant is applied uniformly to all models and therefore the common tangents do not change. \edit{To summarize, we find that some phase transitions e.g.~liquid-FCC are robust in that they exist for any combination of the liquid theory (PY or CS) and lattice free energy calibration constant (1 or $e$), whereas others e.g.~liquid-BCC exist for certain pairs and not for others.} These findings are consistent with previous observations of the robustness of the phase transition to the underlying approximations \cite{colot1986freezing}.

\subsection{Cell theory at low density}
\begin{figure*}
\hspace*{-.62in}\includegraphics[width=1.16\textwidth,trim=0in 0 1.1in 0in, clip]{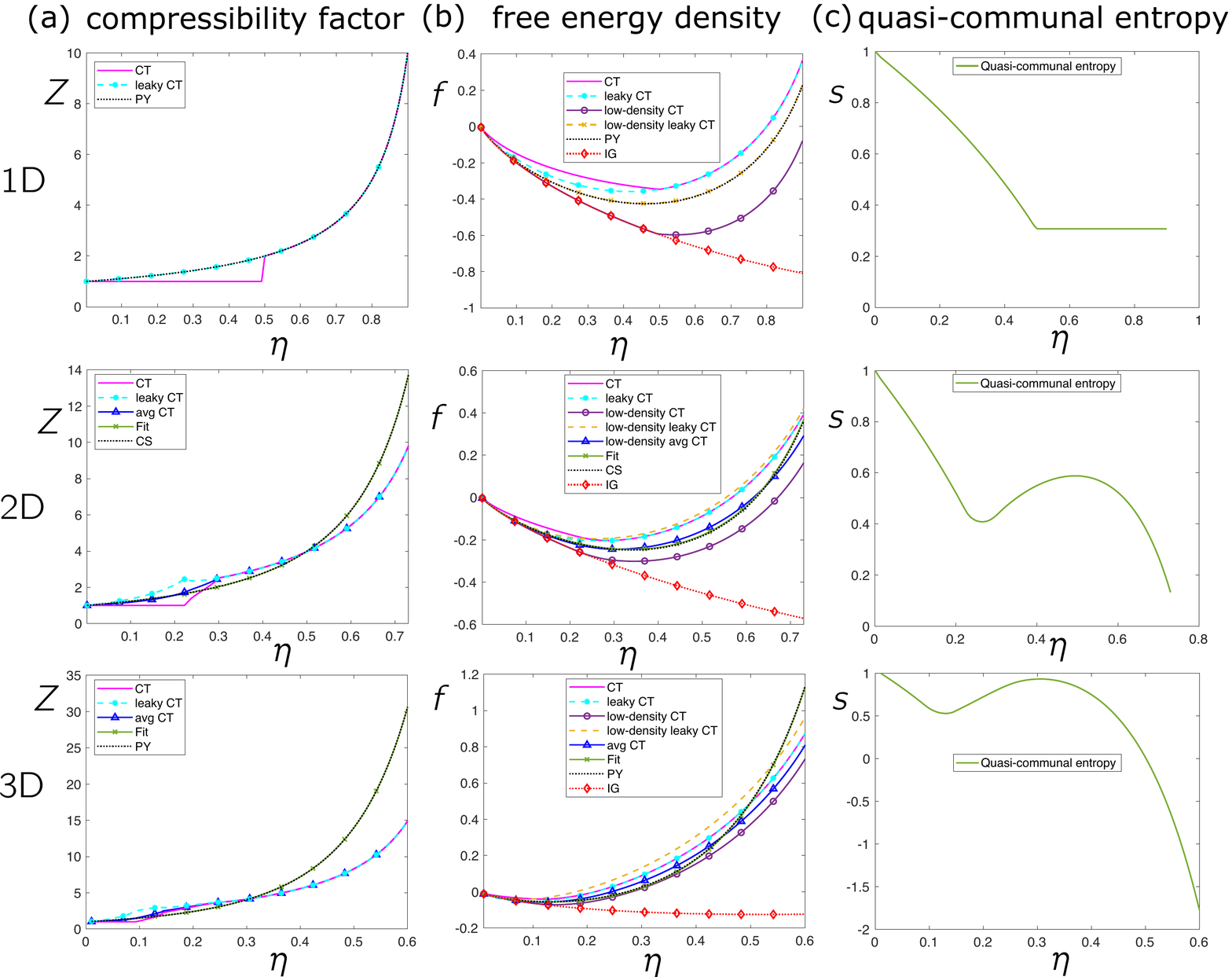}
\caption{\label{fig:comm_ent} Cell theory at low density. First row contains 1D results, second row contains 2D results, and third row contains 3D results. (a) \edit{Compressibility factor}, (b) free energy density, and (c) quasi-communal entropy (the latter two in units of $k_B T$). }
\end{figure*}

Next, to further explore the importance of calibration and to test whether the choice of calibration may affect phase transitions between cell models, we compare versions of the leaky cell model calibrated to low and high density to analogous versions of the standard cell theory (Fig.~\ref{fig:comm_ent}). In the figure legend, those expressions which use the low-density calibration are labeled as ``low-density''; otherwise the high-density calibration is used. 

In one-dimension, we find that the \edit{compressibility factor} predicted by the leaky cell model is exact, and calibrating the free energy to low-density yields a free energy indistinguishable from the exact solution. In two and three dimensions, the leaky cell model is no longer exact. Neither the low-density or high-density calibrations are able to match the empirical equations of state over an appreciable range of packing fractions.

This agreement can be improved somewhat by considering weighted averages of the standard and leaky cell models---weighting the leaky model by $\frac{1}{2}$ and $\frac{1}{3}$ in the 2D and 3D cases, respectively---in a manner similar to the weighting between the hard center and soft center models used in Buehler et al.~\cite{buehler1951free}. However, it is evident that no combination of the cell models provides a strong agreement with the empirical model (unless these weights are varied with $\eta$ and considered as fitting parameters, in which case the cell models lose their straightforward interpretations). This reveals a limitation of the cell theory. No combination of the standard and leaky cell theories can on its own predict the freezing/melting transition in the hard sphere gas. Some additional information is required to capture the behavior at low densities.

\subsection{Quasi-communal entropy}
To address this limitation, inspired by the notion of communal entropy used to capture the error incurred by using the single occupancy assumption at low density \cite{rice1938communal,hoover1968melting}, we augment the free volume by incorporating a quasi-communal entropy $s=s(\eta)$ via:
\begin{equation}
\mathcal{F}'(\eta) = \mathcal{F}^{CT}(\eta)\exp{\left(s(\eta)\right)},
\end{equation}
where $\mathcal{F}^{CT}$ is the free volume in the standard cell theory.

We then fit $s$ to the empirical liquid equation of state using the boundary condition $s(0)=1$ in the low-density limit to recover the additional factor of $e$ per molecule \cite{eyring1937theory}. The quasi-communal entropy $s$ reports on the discrepancy between the standard cell theory and the empirical models, which generally agree with the first several virial coefficients and are therefore a reliable model at low-density. It differs from the notion of communal entropy explored in the literature\cite{kirkwood1950critique,hoover1968melting} in that it is an aggregate measure of the additional entropy in the hard-sphere gas rather than the entropy incurred solely from the single-occupancy assumption.

In the one-dimensional case, the quasi-communal entropy is identical to the communal entropy reported in Hoover and Alder\cite{hoover1966cell}, which used systematic calculations to isolate the effects of the assumptions in cell theory noted by Kirkwood \cite{kirkwood1950critique}. Interestingly, the quasi-communal entropy found in two and three dimensions is a non-monotonic function. As mentioned previously, in addition to single-occupancy the cell theory also assumes independence of neighboring lattice sites, which may be more realistic in the low-density regime \cite{kirkwood1950critique}. Presumably, the non-monotonicity observed in the quasi-communal entropy corresponds in Kirkwood's description of the errors that arise from increased correlations between neighboring cells, which has the effect of increasing the effective free volumes.

To better understand the quasi-communal entropy and how it compares with the actual communal entropy, we revisit \edit{Tonks' gas} in one dimension, for which we can compute the free volume exactly. 
We first review the formulation of Tonks' gas model, following closely Tonks' original paper\cite{tonks1936complete}. Consider a segment of length $\ell$, which hosts $N$ rods of diameter $\sigma$, which glide on the segment without going past each other. We call
\begin{equation}\label{eq:lambda_definition}
\lambda:=\frac{\ell}{N}
\end{equation}
the \emph{specific length}, and
\begin{equation}\label{eq:theta_definition}
\eta:=\frac{\sigma N}{\ell}=\frac{\sigma}{\lambda}	
\end{equation}
the packing fraction as above. In the thermodynamic limit, $N\gg1$, the entropy $S_\mathrm{T}$ of the system is
\begin{equation}\label{eq:entropy_Tonks}
S_\mathrm{T}=k_BTN\left[1+\ln\lambda(1-\eta)\right].
\end{equation}
Letting the force $f$ be defined as
\begin{equation}\label{eq:force_definition}
f:=T\frac{\partial S_\mathrm{T}}{\partial\ell},
\end{equation}	
we arrive at the following expression for the \edit{compressibility factor}
\begin{equation}\label{eq:compressibility}
Z:=\frac{f\ell}{Nk_BT}=\frac{1}{1-\eta}.
\end{equation}

Next, we consider \edit{Tonks' gas} in the context of cell theory, closely following the presentation of Rice\cite{rice1944statistical}. In this formulation, each rod is assumed to be caged in a cell of length $\lambda$ (there are exactly $N$ cells, each confining a single rod). It is a simple matter to derive the entropy $S_\mathrm{R}$ of the system in this case,
\begin{equation}\label{eq:entropy_Rice}
S_\mathrm{R}=k_BTN\ln\lambda(1-\eta),
\end{equation}
so that the \emph{communal} entropy $S_\mathrm{c}$ is simply
\begin{equation}\label{eq:entropy_communal}
S_\mathrm{c}:=S_\mathrm{T}-S_\mathrm{R}=k_BTN
\end{equation}
and the \edit{compressibility factor} $Z$ is still delivered by \eqref{eq:compressibility} (because the communal entropy is constant).

Next, we consider a leaky version of the cell theory above. This was first done by Hoover and Alder \cite{hoover1966cell}, in which the authors considered a first version of leaky cells for Tonks' gas. They allowed the center of mass of each rod (and not just the whole rod) to be confined in a cell and found that, as expected, the communal entropy in this case is smaller than in the case of full confinement. Moreover, the \edit{compressibility factor} is affected, as the communal entropy is no longer constant.

Here, we take a more general view and consider the case in which the center of mass of each rod can get as far as $(\alpha-\frac12)\sigma$ (either way) from the end-points of the cell that bounds it.  Here $0\leqq\alpha\leqq1$ is a fixed \emph{leak parameter}. Formally, the coordinates $x_j$ of the centers of mass of the rods obey the following inequalities,
\begin{equation}
\hspace*{-.14in} \label{eq:coordinate_inequalities} (j-1)\lambda-\left(\alpha-\frac12\right)\sigma\leqq x_j\leqq j\lambda+\left(\alpha-\frac12\right)\sigma,\quad 2\leqq j\leqq N-1.
\end{equation} 
The case $\alpha=0$ corresponds to the model of Rice\cite{rice1944statistical}, whereas the case $\alpha=\frac12$ corresponds to the model of Hoover and Alder\cite{hoover1966cell}. The limiting value $\alpha=1$ is the one where each rod is almost allowed to leave its cell, as it can remain in contact with it at only one point.

\begin{figure}[h!]
	\hspace*{-.2in}\includegraphics[width=0.9\linewidth]{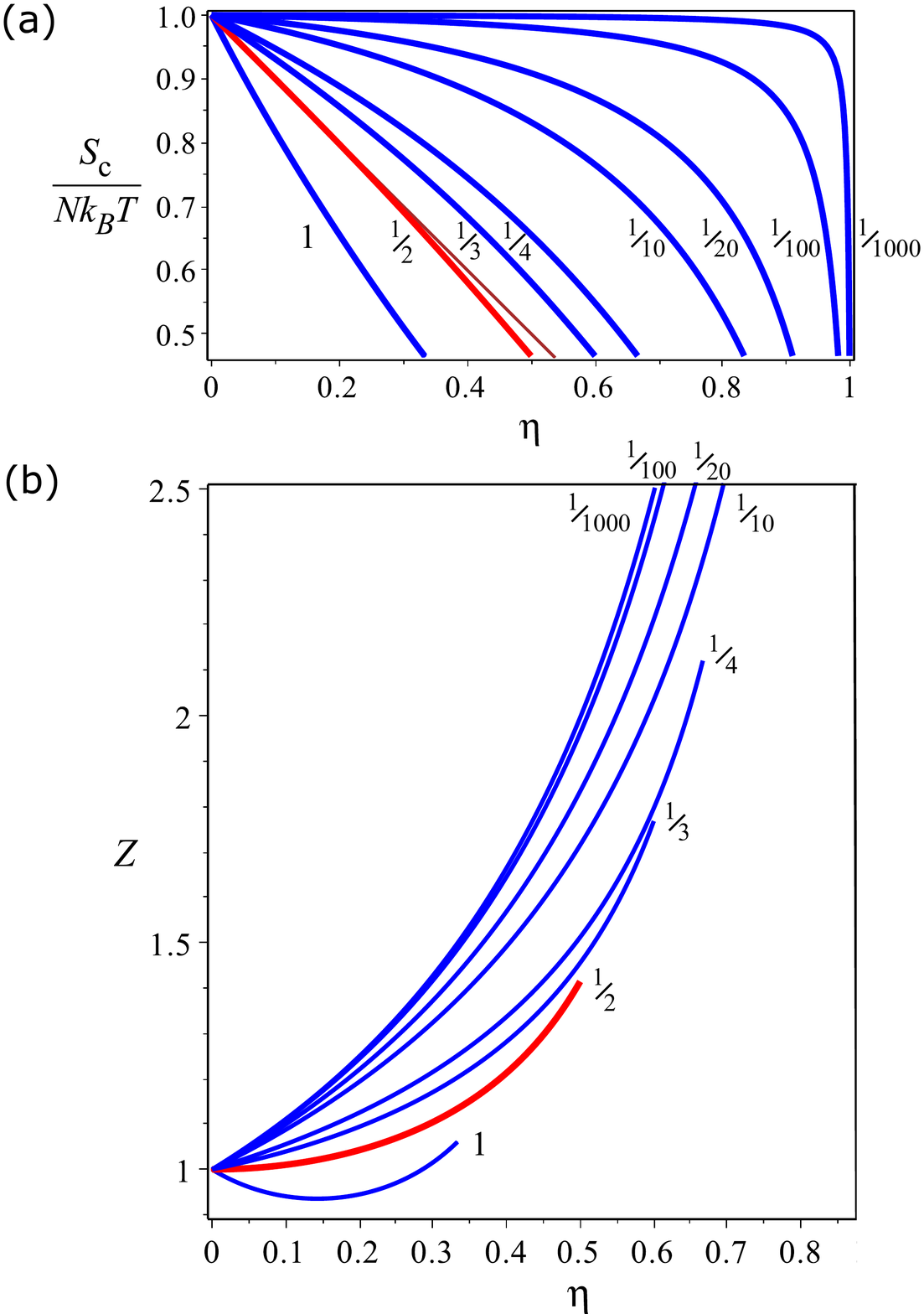}
	\caption{(a) Communal entropy vs.~packing fraction \edit{in the interval \eqref{eq:theta_interval} in which \eqref{eq:entropy_communal_alpha} is valid, plotted for several values of the leak parameter; the red curve corresponds to $\alpha=\oh$. The brown straight line is the graph of the linear function in \eqref{eq:communal_entropy_approximate}.} (b) Corresponding graphs of the \edit{compressibility factor} $Z$ given by \eqref{eq:compressibility_alpha}; the red curve corresponds to $\alpha=\oh$.}
	\label{fig:1d_leaky}
	\end{figure}
As shown in detail in Appendix \ref{app:1D}, it turns out that, at least in the interval
\begin{equation}
\label{eq:theta_interval}
0\leqq\eta\leqq\frac{1}{2\alpha+1},
\end{equation}
the entropy $S$ of the system can be computed exactly (in the thermodynamic limit). It is given by
\begin{equation}
\label{eq:entropy_alpha}
\hspace*{-.14in} S=k_BTN\ln\frac12\lambda\left(1+(2\alpha-1)\eta+\sqrt{[1+(2\alpha-1)\eta]^2-8\alpha^2\eta^2} \right),
\end{equation}
so that the communal entropy becomes
\begin{equation}\label{eq:entropy_communal_alpha}
S_\mathrm{c}=k_BTN\ln\frac{2(1-\eta)\mathrm{e}}{1+(2\alpha-1)\eta+\sqrt{[1+(2\alpha-1)\eta]^2-8\alpha^2\eta^2} }.
\end{equation}
Correspondingly, the \edit{compressibility factor} reads as
\begin{equation}
\label{eq:compressibility_alpha}
Z=\frac{1}{\sqrt{1-2\eta+4\alpha\eta+(1-4\alpha-4\alpha^2)\eta^2}}.
\end{equation}	
Both these functions are plotted in Fig.~\ref{fig:1d_leaky} for different values of $\alpha$, the red graphs corresponding to $\alpha=\frac12$. \edit{A simple computation shows that at the upper limit of validity for \eqref{eq:entropy_communal_alpha}, that is, at $\eta=1/\left(2\alpha+1\right)$,
\begin{equation}
\frac{S_\mathrm{c}}{k_B T N} = 1+\ln\left(\frac{2}{2+\sqrt{2}}\right) \approx 0.47.
\end{equation}
This is precisely the lower bound of the window in which $S_\mathrm{c}/k_B T N$ is plotted in Fig.~\ref{fig:1d_leaky}(a).}

\begin{figure}[h!]
\begin{center}
	\includegraphics[width=.7\linewidth,trim=0in 3.9in 7in 0in, clip]{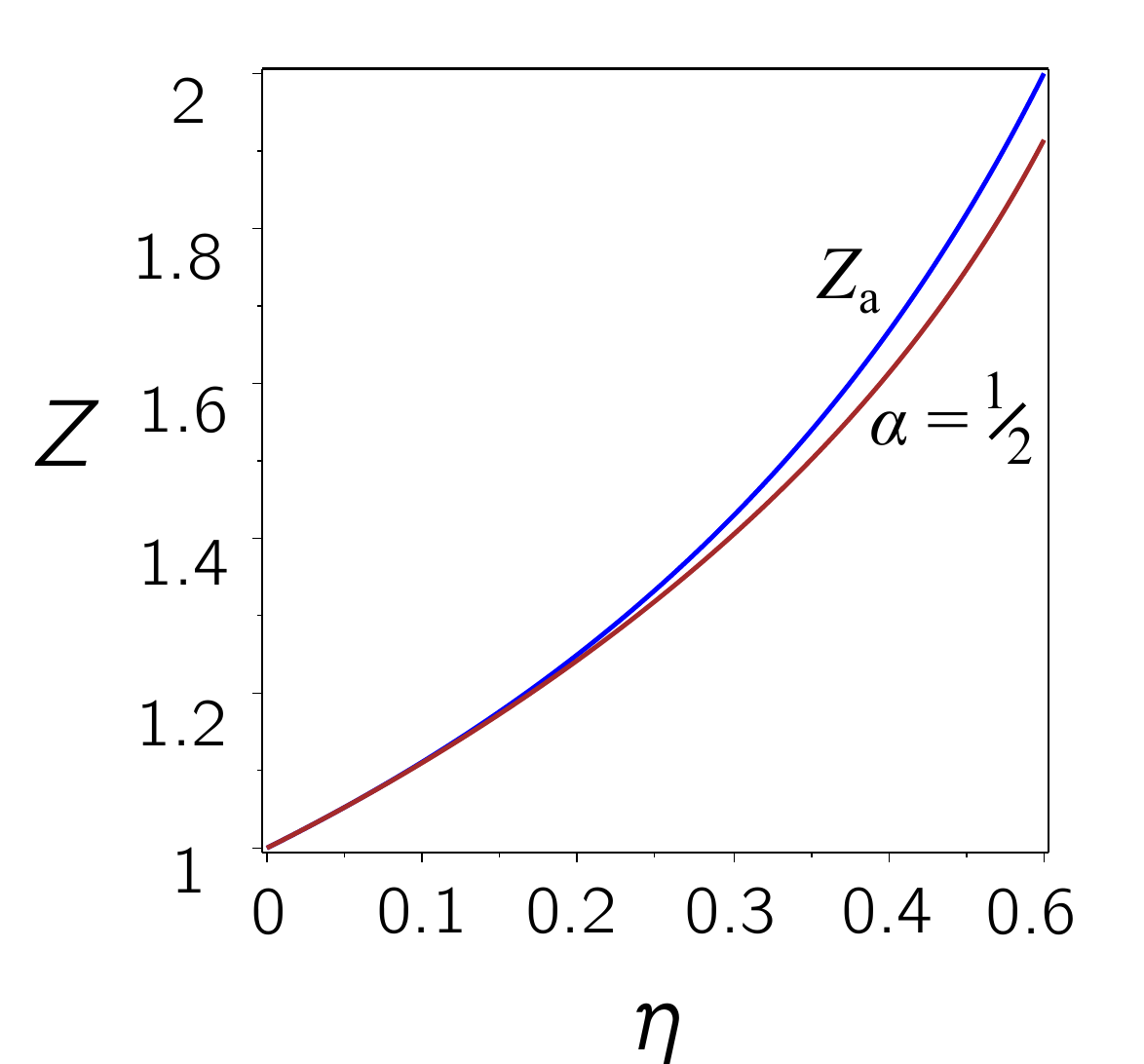}
\end{center}
\vspace{-0.2in}
	\caption{Comparison between the approximate \edit{compressibility factor} $Z_\mathrm{a}$ in \eqref{eq:compressibility_approximate} and the exact \edit{compressibility factor} given by \eqref{eq:compressibility} for $\alpha=\oh$.}
	\label{fig:1d_leaky_comp}
\end{figure}

Approximating the communal entropy for $\alpha=\frac12$ by \edit{the linear function}
\begin{equation}\label{eq:communal_entropy_approximate}
S_\mathrm{c}\approx k_BTN(1-\eta),
\end{equation}
as suggested by the comparison between red and brown graphs in Fig.~\ref{fig:1d_leaky}(a), one obtains the following approximate form for the \edit{compressibility factor},
\begin{equation}
\label{eq:compressibility_approximate}
Z_\mathrm{a}=\frac{1}{\sqrt{1-2\eta^2}}+\eta,
\end{equation}
which is plotted (in brown) \edit{in Fig.~\ref{fig:1d_leaky_comp}} against the exact \edit{compressibility factor} from \eqref{eq:compressibility} (in blue).

The calculation above makes clear some of the ways in which the quasi-communal entropy differs from the bona fide communal entropy. In particular, the true communal entropy is a monotonic function bounded between 0 and 1 for any value of the leak parameter. The fact that the quasi-communal entropy does not satisfy these criteria reveals that it cannot be interpreted as the error incurred by the single-occupancy assumption. Rather, it is properly interpretated as a net error that includes not only the single-occupancy assumption but also errors due to other approximations e.g.~that the free volumes of neighboring cells are uncorrelated.

\section{Discussion}
In this work, we extend the classical cell theory into an intermediate leaky regime in which the local neighborhood of spheres expands, with consequences on the resulting free volume. The relevant leaky packing fractions are $\eta_l\approx0.26$ for the FCC lattice, $\eta_l\approx0.29$ for the SC lattice, and $\eta_l\approx0.37$ for the BCC lattice. Below this leaky cell fraction and above the corresponding percolation thresholds, spheres remain locally caged by their neighbors but the cage involves additional neighbors. Interestingly, the effect of the leaky cell model is not uniform for all three of these lattices. In the FCC and SC lattices, it leads to additional free volume, whereas in the BCC lattice it leads to a decrease. This is a direct consequence of the geometry of the lattice, as we show explicitly through exact calculations of the free volume through the leaky regime.

In the leaky cell model explored here, we allow for an intermediate regime in which spheres may escape their Voronoi cells while they remain sterically trapped within an expanded cage. By expressing the free energy in terms of the leaky cell volume, which may be calculated exactly, we apply the leaky cell model to various lattice arrangments, and predict phase transitions between lattices in the leaky regime.

We have identified two interlattice phase transitions within this intermediate leaky regime. There are SC-FCC and SC-BCC phase transitions around $\eta=0.22$. In addition, there is a quasi intralattice phase transition around $\eta=0.3$ within the SC lattice, which arises because of a discontinuity in the free energy density. These phase transitions indicate the possibility that different lattice structures may coexist within \edit{crystalline} materials. Such cubic-cubic phase transitions have previously been observed experimentally in fullerene crystals \cite{david1992structural,wochner1997x}, and the leaky model may be an apt description of such systems. However, it has been shown previously\cite{salsburg1962equation} that not all lattice configurations are stable, and that stability to shear requires certain topological constraints on the number of contacts to be satisfied. Therefore, physical forces beyond those of pure hard sphere models may be required to realize some of the lattices discussed in this work.

An early critique of the cell theory was that the mean-field assumptions underlying the model, such as the single occupancy of cells and the independence of neighboring free volumes, were not always clearly specified \cite{kirkwood1950critique,wood1952note,hoover1966cell}. In particular, depending on the assumptions made, the resulting free energy may differ by an additive constant. In this work we consider the effect of such translations in the free energy in the context of the leaky cell model. We have found that some of the phase transitions found are sensitive to this calibration, particularly the freezing/melting transition in the BCC lattice, in which the existence of a phase transition is sensitive to the choice of liquid model and calibration. On the other hand, we find that predicting the known freezing/melting transition in a hard sphere gas requires information beyond the cell theory, such as an empirical liquid model (e.g. Carnahan-Starling): regardless of whether it is calibrated to low or high density or some simple combination, when the liquid model is based purely on a cell theory we find no freezing transition reminiscent of the hard sphere gas.

Finally, we comment on possible extensions of this model. Although hard sphere models are unable to capture the behavior of systems such as liquid crystals in which the particles are rod-shaped, this work represents a step along the path to understanding how cell theories based on free volume may be used in non-spherical systems, as has recently been done in the context of predicting the glass transition \cite{gujrati2021}.
Although it may not be possible to obtain exact formulas in these more complicated settings, phase transitions could be identified by using numerical methods to approximate the free energy, as has been done previously using Monte Carlo simulation \cite{barroso2002solid,schilling2009computing}. The formulas presented here are strictly valid on lattices with periodic boundary conditions. In the case of a finite domain with walls, it may be possible to calculate finite size corrections for the leaky cell model as has been done for other lattice structures \cite{polson2000finite}.

\begin{acknowledgments}
We thank Zachary M.~Geballe for reading an early draft of this manuscript and providing helpful feedback. T.G.F.~acknowledges support from National Science Foundation grant DMS-1913093. P.P-M.~acknowledges support from the Office of Naval Research (ONR N00014-18-1-2624). J.M.T.~has been partially supported by the Basque Government through the BERC 2018-2021 program; and by Spanish Ministry of Economy and Competitiveness MINECO through BCAM Severo Ochoa excellence accreditation SEV-2017-0718 and through project MTM2017-82184-R funded by (AEI/FEDER, UE) and acronym ``DESFLU". Finally, we wish to acknowledge the support of the Institute for Mathematics and its Applications (IMA), where this work was initiated during the 2018 program on ``Multiscale Mathematics and Computing in Science and Engineering''. 
\end{acknowledgments}

\section*{Data Availability}
The code used to perform the calculations presented here is available on GitHub: \url{https://github.com/thomasgfai/LeakyCellModel}.
\appendix

\section{\label{app:exact} Exact free volumes on 3D lattices}
To calculate exact free volumes accessible to spheres, we use the following geometric idea. From the volume of the void space made up of a central sphere and its neighbors, we first subtract the volumes of the exclusion spheres of radius $2R$ with their corresponding solid angles. If the exclusion spheres overlap, we must correct for this double-counting by adding the volumes of double intersections between exclusion spheres. If the double intersections overlap, we must subtract a triple intersection, and so on. Because there are general formulas for the volumes of spherical double, triple, and quadruple intersections \cite{gibson1987exact,gibson1987volume}, this method leads to an exact formula for the free volume that can in be applied to any lattice.

\subsection{FCC/HCP lattice}
The FCC lattice involves planar layers of hexagonally-arranged spheres. The surrounding void space can be partitioned into $N_t=8$ tetrahedra and $N_o=6$ octahedra. As the packing fraction increases, a centrally-located sphere becomes caged first by triangular faces (with 18 neighboring spheres) and subsequently by octahedral midplanes (with 12 neighboring spheres). See Fig.~\ref{fig:loc_pack}(a)--(b). Note that the cage in the left column of Fig.~\ref{fig:loc_pack} is always a subset of the corresponding cage in the right column. For example, for the FCC lattice illustrated in the first row of  Fig.~\ref{fig:loc_pack}, the cage in the right column is obtained by adding outward-pointing right square pyramids to the cage in the left column.

We express the free volume in terms of the edge length $a$. It is straightforward to write these in terms of the Voronoi cell volume, as the edge length $a$ is related to the Voronoi cell volume $v$ through the relation
\begin{equation}
v = \frac{N_t}{4}V_t+\frac{N_o}{6}V_o,
\end{equation}
where the numerical factors $4$ and $6$ come from the number of vertices of a tetrahedron and octahedron and where $V_t=a^3/\left(6\sqrt{2}\right)$ and $V_o=\sqrt{2}a^3/3$ are their respective volumes. It follows that
\begin{equation}
v(a) = \frac{a^3}{\sqrt{2}}.
\end{equation}
We use the notation $V_s:=\frac{4}{3}\pi (2R)^3$ to denote the volume of a sphere of radius $2R$. The free area $\mathcal{F}(v)$ on the FCC lattice satisfies
\begin{equation}
\hspace{-.05in}\mathcal{F}(v) =\begin{cases}
  0, \qquad &a < 2R\\
  \\[-.2in]
  N_t\left(V_t-3V_s \frac{\Omega_t}{4\pi}+3 V_{2I}(a) \frac{\alpha_t}{2\pi}\right) \qquad & \\
\quad +N_o\left(V_o/2-2V_s \frac{\Omega_o}{4\pi}+2 V_{2I}(a) \frac{\alpha_o}{2\pi} \right.\\
\left. \quad +V_{2I}(\sqrt{2}a)-2V_{3I}^\text{rt}(a)+\oh V_{4I}(a) \right) \\
\quad -4V_{3I}^\text{eq}(a),  \qquad &2R < a < 2\sqrt{2} R  \\
   \\[-.2in]
     N_t\left(V_t-3V_s \frac{\Omega_t}{4\pi}+3 V_{2I}(a) \frac{\alpha_t}{2\pi}\right) \qquad & \\
\quad +N_o\left(V_o-5V_s \frac{\Omega_o}{4\pi}+8 V_{2I}(a) \frac{\alpha_o}{2\pi} \right) \\
\quad -16V_{3I}^\text{eq}(a),  \qquad &2R < a < 2\sqrt{3} R  \\
   \\[-.2in]
  N_t\left(V_t-3V_s \frac{\Omega_t}{4\pi}+3 V_{2I}(a) \frac{\alpha_t}{2\pi}\right) \qquad & \\
\quad +N_o\left(V_o-5V_s \frac{\Omega_o}{4\pi}+8 V_{2I}(a) \frac{\alpha_o}{2\pi}\right), \qquad &2\sqrt{3} R<a < 4R\\
    \\[-.2in]
  N_t\left(V_t-3V_s \frac{\Omega_t}{4\pi}\right) \qquad & \\
\quad +N_o\left(V_o-5V_s \frac{\Omega_o}{4\pi}\right), \qquad & 4R<a,
\end{cases}
\label{eq:fv_hcp} 
\end{equation}
where $\Omega_t=\cos^{-1}(23/27)$, $\Omega_o=4 \sin^{-1}(1/3)$ are the solid angles subtended by the vertices of regular tetrahedra and octahedra and $\alpha_t=\cos^{-1}(1/3)$ and $\alpha_o=\cos^{-1}(-1/3)$ are the corresponding dihedral angles. 
Note that beyond the percolation threshold, at which $a > 2\sqrt{3} R$, the sphere can escape through tetrahedral faces. However, we do not count the volume outside of the cage so that free volume remains an intensive quantity and the system size $L$ does not enter the formula.

In addition to the double intersection volume $V_{2I}(a)$, in Eq.~\eqref{eq:fv_hcp} we have used the following notation: $V_{3I}^\text{eq}(a)$ is the intersection volume of three spheres at the vertices of an equilateral triangle of edge-length $a$, $V_{3I}^\text{rt}(a)$ is the intersection volume of three spheres along a right triangle at vertices of a square of edge-length $a$, and finally $V_{4I}(a)$ is the quadruple intersection volume of four spheres at the vertices of a square of edge-length $a$.

The volume of intersection between two spheres of radius $2R$ separated by a distance $a$ is given by
\begin{equation}
V_{2I}(a) = \frac{\pi}{12}\left(4(2R)+a\right)\left(2(2R)-a\right)^2.
\end{equation}
General formulas for volumes of higher-order intersections are given in Gibson and Scheraga \cite{gibson1987exact,gibson1987volume}. In particular,
\begin{align}
V_{3I}^\text{eq}(a) &= \frac{a^2w^\text{eq}}{6}-\frac{3a}{2}\left(2(2R)^2-\frac{a^2}{6}\right)\tan^{-1}\left(\frac{2w^\text{eq}}{a}\right) \notag\\
&\quad+4(2R)^3\tan^{-1}\left(\frac{w^\text{eq}}{2R}\right)
\end{align}
where
\[w^\text{eq} = \sqrt{3(2R)^2-a^2} \quad \qquad \qquad \qquad \qquad \qquad\]
and
\begin{align}
V_{3I}^\text{rt}(a) &= \frac{a^2w^\text{rt}}{6}-a\left(2(2R)^2-\frac{a^2}{6}\right)\tan^{-1}\left(\frac{w^\text{rt}}{a}\right) \notag\\
  &\quad-\frac{a\sqrt{2}}{2}\left(2(2R)^2-\frac{a^2}{3}\right)\frac{\pi}{2} \\
&\quad +\frac{4(2R)^3}{3}\left(2\tan^{-1}\left(\frac{w^\text{rt}}{2(2R)}\right)+\frac{\pi}{2}\right) \notag
\end{align}
where
\[w^\text{rt} = \sqrt{4(2R)^2-2a^2}. \quad \qquad \qquad \qquad \qquad\]
As explained in \citen{gibson1987exact}, the quadruple intersection volume may be computed simply by	
\begin{equation}
V_{4I}(a)=2V_{3I}^\text{rt}(a)-V_{2I}(\sqrt{2}a).
\end{equation}

\subsection{BCC lattice}
In the BCC lattice the percolation threshold occurs at $\eta^\text{BCC}_\text{p} \approx 0.19$, at which spheres are caged by 14 neighbors, located both at vertices of a cube as well as across cubic faces. The void space is contained within 6 octahedra (regular square pyramids mirrored at the base). At the leaky cell transition $\eta^\text{BCC}_\text{l} \approx 0.19$, spheres can no longer escape through the square octahedral midplanes, so that the cage is formed by the eight neighbors on cube vertices. Unlike the other lattices, the leaky cell model in the BCC lattice leads to \textit{lower} free volumes since neighbors across cubic faces occlude volume via their exclusion spheres leaking into the Wigner-Seitz cell.

Upon making the identification $a(v) = \sqrt[3]{2v}$, $\mathcal{F}(v)$ may be written as
\begin{equation}
\hspace{-.45in} \mathcal{F}(v)=
\begin{cases}
  0, &a < \frac{4}{\sqrt{3}R}\\
  \\[-.2in]
  a^3-V_s+3V_{2I}(a)+6V_{2I}(\sqrt{2}a) &\\
  \quad-12V_{3I}^\text{eq}(a)+3V_{4I}(a), &\frac{4}{\sqrt{3}R} < a < \frac{8}{3} R \\
     \\[-.2in]
  N_oV_o-N_o\left(4V_s\frac{2\Omega_b}{4\pi}+V_s\frac{\Omega_a}{4\pi}\right. &\\
  \quad\left.+4V_{2I}(a)\frac{2\alpha_b}{2\pi} +4V_{2I}(\frac{\sqrt{3}}{2}a)\frac{\alpha_a}{2\pi}\right. &\\
  \quad\left.+2V_{2I}(\sqrt{2}a)-2V_{3I}^\text{iso}(a,\frac{\sqrt{3}}{2}a)\right. &\\
  \quad\left.-4V_{3I}^\text{rt}(a)-2\widetilde{V}_{3I}^\text{iso}(\sqrt{2}a,\frac{\sqrt{3}}{2}a)\right. &\\
  \quad\left.+4\widetilde{V}^\text{pyr}_{4I}(a,\frac{\sqrt{3}}{2}a)\right) &\\
  \quad\left.+4V^\text{sq}_{4I}(a)-4\widetilde{V}^\text{pyr}_{5I}(a,\frac{\sqrt{3}}{2}a)\right), &\frac{8}{3} R < a < 2\sqrt{2} R\\
     \\[-.2in]
  N_oV_o-N_o\left(4V_s\frac{2\Omega_b}{4\pi}+V_s\frac{\Omega_a}{4\pi} \right. &\\
  \quad\left.+4V_{2I}(a)\frac{2\alpha_b}{2\pi}+4V_{2I}(\frac{\sqrt{3}}{2}a)\frac{\alpha_a}{2\pi}\right. &\\
  \quad\left.-2V_{3I}^\text{iso}(a,\frac{\sqrt{3}}{2}a)\right), &2\sqrt{2} R < a < \frac{8\sqrt{2}}{3} R\\
   \\[-.2in]
  N_oV_o-N_o\left(4V_s\frac{2\Omega_b}{4\pi}+V_s\frac{\Omega_a}{4\pi} \right. &\\
  \quad\left.+4V_{2I}(a)\frac{2\alpha_b}{2\pi}+4V_{2I}(\frac{\sqrt{3}}{2}a)\frac{\alpha_a}{2\pi}\right),  & \frac{8\sqrt{2}}{3} R < a < 4 R\\
  \\[-.2in]
  N_oV_o-N_o\left(4V_s\frac{2\Omega_b}{4\pi}+V_s\frac{\Omega_a}{4\pi} \right. &\\
  \quad\left.+4V_{2I}(\frac{\sqrt{3}}{2}a)\frac{\alpha_a}{2\pi}\right),  &4 R < a < \frac{8}{\sqrt{3}} R\\
      \\[-.2in]
  N_oV_o-N_o\left(4V_s\frac{2\Omega_b}{4\pi}+V_s\frac{\Omega_a}{4\pi} \right), &\frac{8}{\sqrt{3}} R<a.
\end{cases}
\label{eq:fv_bcc} 
\end{equation}
Here, $\Omega_a=2\pi/3$ and $\Omega_b=\pi/6$ are the solid angles subtended by the apex and base, respectively, of the regular square pyramid, and $\alpha_a=\cos^{-1}(-\oh)$ and $\alpha_b=\pi/4$ are the corresponding dihedral angles. Further, $V_{3I}^\text{iso}(x,y)$ is the triple intersection between three spheres at vertices of an isosceles triangle with edge lengths $x$, $y$, and $y$. To calculate $V_{3I}^\text{iso}$, we use the general formula from \citen{gibson1987exact}:
\begin{align}
V_{3I}^\text{iso}(x,y) &= \frac{a^2w^\text{rt}}{6}-a\left(2(2R)^2-\frac{a^2}{6}\right)\tan^{-1}\left(\frac{w^\text{rt}}{a}\right) \notag\\
  &\quad-\frac{a\sqrt{2}}{2}\left(2(2R)^2-\frac{a^2}{3}\right)\frac{\pi}{2} \\
& \quad+\frac{4(2R)^3}{3}\left(2\tan^{-1}\left(\frac{w^\text{rt}}{2(2R)}\right)+\frac{\pi}{2}\right) \notag
\end{align}
where
\[w^\text{rt} = \sqrt{4(2R)^2-2a^2}. \quad \qquad \qquad \qquad \qquad\]
The remaining intersections appearing in \eqref{eq:fv_bcc} simplify upon inspection; one of the intersections does not exclude any additional volume, and all the intersections marked with tildes are equal to lower-order intersections. In particular, the triple intersection $\widetilde{V}_{3I}^\text{iso}(\sqrt{2}a,\frac{\sqrt{3}}{2}a)$ formed between the spheres at apex of the regular right pyramid and vertices along a diagonal at its base is in fact a double intersection, i.e.
\[\widetilde{V}_{3I}^\text{iso}(\sqrt{2}a,\frac{\sqrt{3}}{2}a) = V_{2I}(\sqrt{2}a).\]
Similarly, the quadruple intersection $\widetilde{V}^\text{pyr}_{4I}(a,\frac{\sqrt{3}}{2}a)$ formed between the sphere at the apex of the regular square pyramid and three spheres at its base reduces to a triple intersection, so that
\[\widetilde{V}^\text{pyr}_{4I}(a,\frac{\sqrt{3}}{2}a) = V_{3I}^\text{rt}(a).\]
Finally, the quintuple intersection $\widetilde{V}^\text{pyr}_{5I}(a,\frac{\sqrt{3}}{2}a)$ formed between the sphere at the apex of the regular square pyramid and all four spheres at its base is simply a quadruple intersection, i.e.
\[\widetilde{V}^\text{pyr}_{5I}(a,\frac{\sqrt{3}}{2}a) = V^\text{sq}_{4I}(a).\]

\subsection{SC lattice}
In addition to the percolation threshold at $\eta^\text{SC}_\text{p} \approx 0.19$, at which diagonal neighbors on square faces begin to intersect and spheres becomes caged by 26 neighbors, there is a transition in the cubic lattice at $\eta^\text{SC}_\text{l} \approx 0.29$ beyond which the sphere can no longer escape through triangular faces, so that opposite corners of cubes are lost and the number of neighbors decreases to 6.

Upon making the identification $a(v) = \sqrt[3]{v}$, $\mathcal{F}(v)$ may be written as
\begin{equation}
\hspace{-.25in}  \mathcal{F}(v)=
\begin{cases}
 0, \qquad &a < 2R\\
 \frac{4}{3}a^3-6 V_s \frac{\Omega_o}{4\pi}+12 V_{2I}(\sqrt{2}a)\frac{\alpha_o}{2\pi} \\
  \quad -4V_{3I}^\text{eq}(\sqrt{2}a) , \qquad &2R < a < \sqrt{6}R\\
  8a^3-7V_s+18 V_{2I}(a)+36V_{2I}(\sqrt{2}a) \\
           \quad -60V_{3I}^\text{rt}(a)+12V_{4I}(a), \qquad &\sqrt{6}R < a < 2\sqrt{2}R\\
  8a^3-7V_s+18 V_{2I}(a), \qquad &2\sqrt{2}R < a < 4R\\
  8a^3-7V_s, \qquad & 4R<a,
\end{cases}
\end{equation}
where as before $\Omega_o=4 \sin^{-1}(1/3)$ is the solid angle subtended by the vertices of a regular octahedron, $\alpha_o=\cos^{-1}(-1/3)$ is the corresponding dihedral angle, and the quadruple intersection may be computed by $V_{4I}(a)=2V_{3I}^\text{rt}(a)-V_{2I}(\sqrt{2}a)$. Note that there is a discontinuity in the second derivative of $\mathcal{F}(v)$ at the leaky cell transition.

\section{\label{app:approx} Asymptotic approximation of the free volume}
\begin{figure}\begin{center}
\includegraphics[width=0.5\textwidth]{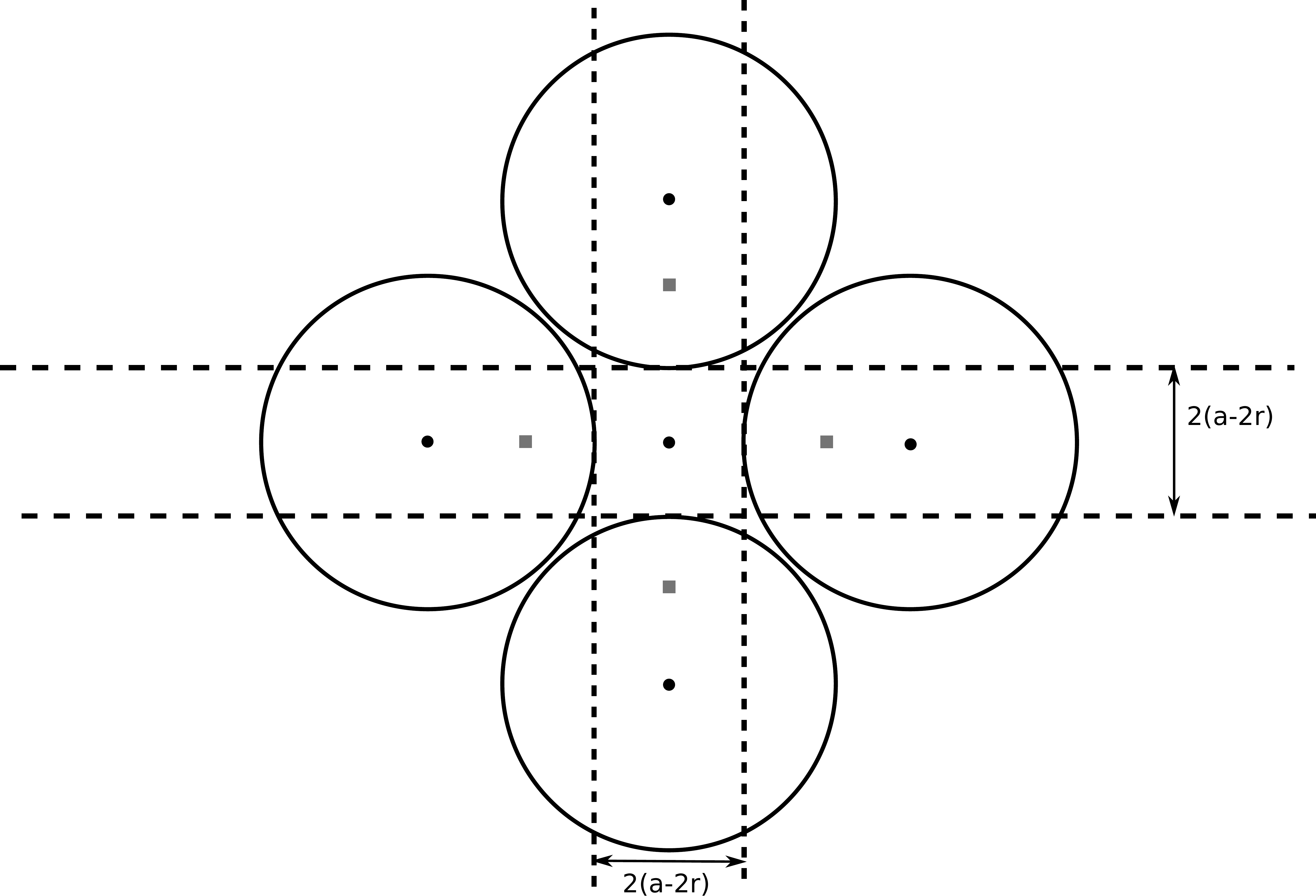}
\caption{Schematic of the polygonal approximation}\label{figSchema}
\end{center}
\end{figure}

Consider a general lattice, which is jammed at its densest packing $\eta_\text{cp}$. We illustrate the situation in Fig.~\ref{figSchema} using a square lattice in 2D, where we show the centers of mass of the centre particle and its neighbors (black dots), and the exclusion regions due to its nearest neighbors (black circles). We approximate the void space by a polygon. That is, we approximate the free volume of a particle at $0$ with nearest neighbors at $a\mb{e}_i$, for $a$ a lattice spacing parameter and $\mb{e}_i$ unit vectors by replacing the exclusion regions $\norm{a\mb{e}_i-\mb{x}}<2r$ with half-spaces, $\mb{e}_i\cdot \mb{x} >a-2r$.  In the schematic Fig.~\ref{figSchema}, the boundaries of these half spaces are given by the dashed lines, and the polygonal cavity corresponds to the square contained within. First of all, we note that this polygonal cavity will simply be a uniform scaling of the Voronoi cell of the densest packing. We can see this because the bounding hyperplanes correspond to the shared faces of the Voronoi cells of particles at $0$ and $2(a-2r)\mb{e}_i$, which is simply a rescaled version of our lattice. These ``virtual point particles" bounding the Voronoi cells are given by grey squares in Fig.~\ref{figSchema}. The characteristic width of this Voroni cell is thus $2(a-2r)$, in comparison to $2r$ at the densest packing. Thus the scale factor between the two Voronoi cells is given by $2\left(\frac{a}{2r}-1\right)$, and thus the volume $\mathcal{F}$ of the polygonal cavity is given by 
\begin{equation}
\mathcal{F}(a)\approx 2^d\left(\frac{a}{2r}-1\right)^dv_\text{cp}=\frac{2^d\left(\frac{a}{2r}-1\right)^d}{\rho_\text{cp}},
\end{equation}
where $v_\text{cp}$ is the volume of the Voronoi cell in the densest packing, and $\rho_\text{cp}$ is the number density at densest packing, related by $\rho_\text{cp}^{-1}=v_\text{cp}$. The number density of a lattice with spacing $a$ is readily given by 
\begin{equation}
\rho=\left(\frac{2r}{a}\right)^{d}\rho_\text{cp},
\end{equation}
by noting that $\frac{a}{2r}$ is the scale factor between the densest packing and the lattice with parameter $a$. We may substitute this into our previous expression to give 
\begin{equation}
\mathcal{F}(\rho)\approx \frac{2^d\left(\sqrt[d]{\frac{\rho_\text{cp}}{\rho}}-1\right)^d}{\rho_\text{cp}}.
\end{equation}
Remarkably, this only depends on the geometry of the lattice via the densest packing parameter. 

It is worth noting that as both $\mathcal{F}$ and this approximation, $\mathcal{F}_\text{approx}$ converge to $0$ as $\rho\nearrow\rho_\text{cp}$, this means that 
\begin{equation}
\lim\limits_{\rho\to\rho_\text{cp}^-}\frac{\mathcal{F}(\rho)}{\mathcal{F}_\text{approx}(\rho)}=\lim\limits_{\rho\to\rho_\text{cp}^-}\frac{\mathcal{F}'(\rho)}{\mathcal{F}_\text{approx}'(\rho)},
\end{equation}
by l'H\^{o}pital's rule, so that if $v_{ap}$ is a good approximation to 0\textsuperscript{th}-order (in the sense that the left-hand limit in the previous equation equals $1$), it is automatically a good approximation to 1\textsuperscript{st}-order (in the sense the right-hand limit equals $1$). It is precisely this approximation which is used in Fig.~\ref{fig:comp} of the Main Text.

\section{Exact free volumes on 2D lattices}
\label{app:2D}
In this appendix, we provide the analytic free volume formulas for lattice packings in 2D using the same geometric argument applied in 3D.
\subsection{Hexagonal lattice}
For a hexagonal lattice of disks of radius $R$, the percolation threshold $\eta^\text{HE}_\text{p}$ is reached when the lattice edge length $a$ satisfies $a = 4R$, corresponding to $\eta^\text{HE}_\text{p} \approx 0.23$. The close-packing limit $\eta^\text{HE}_\text{cp}$ is attained when $a = 2R$, corresponding to $\eta^\text{HE}_\text{cp} \approx 0.91$, at which point the disk is in contact with its neighbor so that it has no free volume to explore. Unlike any of the lattices investigated in the 3D case, there is no leaky regime in the hexagonal 2D lattice; disks escape from their unit cells at the percolation threshold, and their is no emergence of an expanded cage that involves additional neighbors. The critical packing fractions of the 2D cell theory are summarized in Table~\ref{tab:2Dfractions}.
\begin{table}
\caption{\label{tab:2Dfractions}Percolation ($\eta_\text{p}$), leaky ($\eta_\text{l}$), and close-packed ($\eta_\text{cp}$) packing fractions in the hexagonal and square 2D lattices.}
\begin{ruledtabular}
\begin{tabular}{lccr}
2D Lattice&$\eta_\text{p}$&$\eta_\text{l}$&$\eta_\text{cp}$\\[0.4em] \hline\\[-0.8em]
hexagonal (HE)&$\frac{\pi}{8\sqrt{3}}\approx 0.23$&N/A&$\frac{\pi}{2\sqrt{3}}\approx 0.91$ \\[0.8em]
square (SQ)&$\pi/16 \approx 0.20$&$\frac{\pi}{8}\approx 0.39$&$\frac{\pi}{4}\approx 0.79$
\end{tabular}
\end{ruledtabular}
\end{table}

Given a hexagonal lattice of $N$ disks in a 2D $L\times L$ domain, using the identification $a(v) = \sqrt{\frac{2v}{\sqrt{3}}}$ that relates edge lengths and local volumes in a hexagonal lattice, the free volume satisfies
\begin{equation}
 \mathcal{F}(v)=
\begin{cases}
 0, \qquad &a < 2R\\
  3v-2V_s+3V_{2I}(a), \qquad &2R < a < 4R\\
   3v-2V_s, \qquad &4R<a,
\end{cases}
\label{eq:fv_hex}
\end{equation}
where $V_s=\pi (2R)^2$ is the area of a disk of radius $2R$ and where the area of intersection of two such disks separated by a distance $a$ is given by
\[ V_{2I}(a) := 2(2R)^2\cos^{-1}\left(\frac{a}{4R}\right)-\frac{a}{2}\sqrt{4(2R)^2-a^2}.\]

\subsection{Square lattice}
Next, we provide the corresponding formula for the free area in a two-dimensional square lattice.
The percolation transition $\eta^\text{SQ}_\text{p}$ at which $a = 4R$ satisfies $\eta^\text{SQ}_\text{p} \approx 0.20$. 
The close-packing fraction $\eta^\text{SQ}_\text{cp}$ at which $a = 2R$ occurs at $\eta^\text{SQ}_\text{cp} \approx 0.79$. 
In addition, there is a leaky cell transition $\eta^\text{SQ}_\text{l} \approx 0.39$ below which the number of neighbors caging each disk expands from 8 to 4. 
Above the leaky cell fraction, which corresponds to edge lengths of $a=2\sqrt{2}R$, disks becomes caged by their orthogonal neighbors (i.e.~the corners are lost).

Upon making the identification $a(v) = \sqrt{v}$ for a square lattice, $\mathcal{F}(v)$ may be written as
\begin{equation}
 \mathcal{F}(v)=
\begin{cases}
  0, \qquad &a < 2R\\
  2v-V_s+2V_{2I}(\sqrt{2}a), \qquad &2R < a < 2\sqrt{2}R\\
  4v-3V_s+4V_{2I}(a), \qquad &2\sqrt{2}R < a < 4R\\
   4v-3V_s, \qquad &a > 4R.
\end{cases}
\end{equation}
Note that, in contrast to the smooth free volume derived in the hexagonal case, there is a discontinuity in the second derivative of $\mathcal{F}(v)$ at the leaky cell transition.

\begin{figure}[hptb]
\centering
	 \includegraphics[trim=.45in 5.5in 0 0, clip,width=3.5 in]{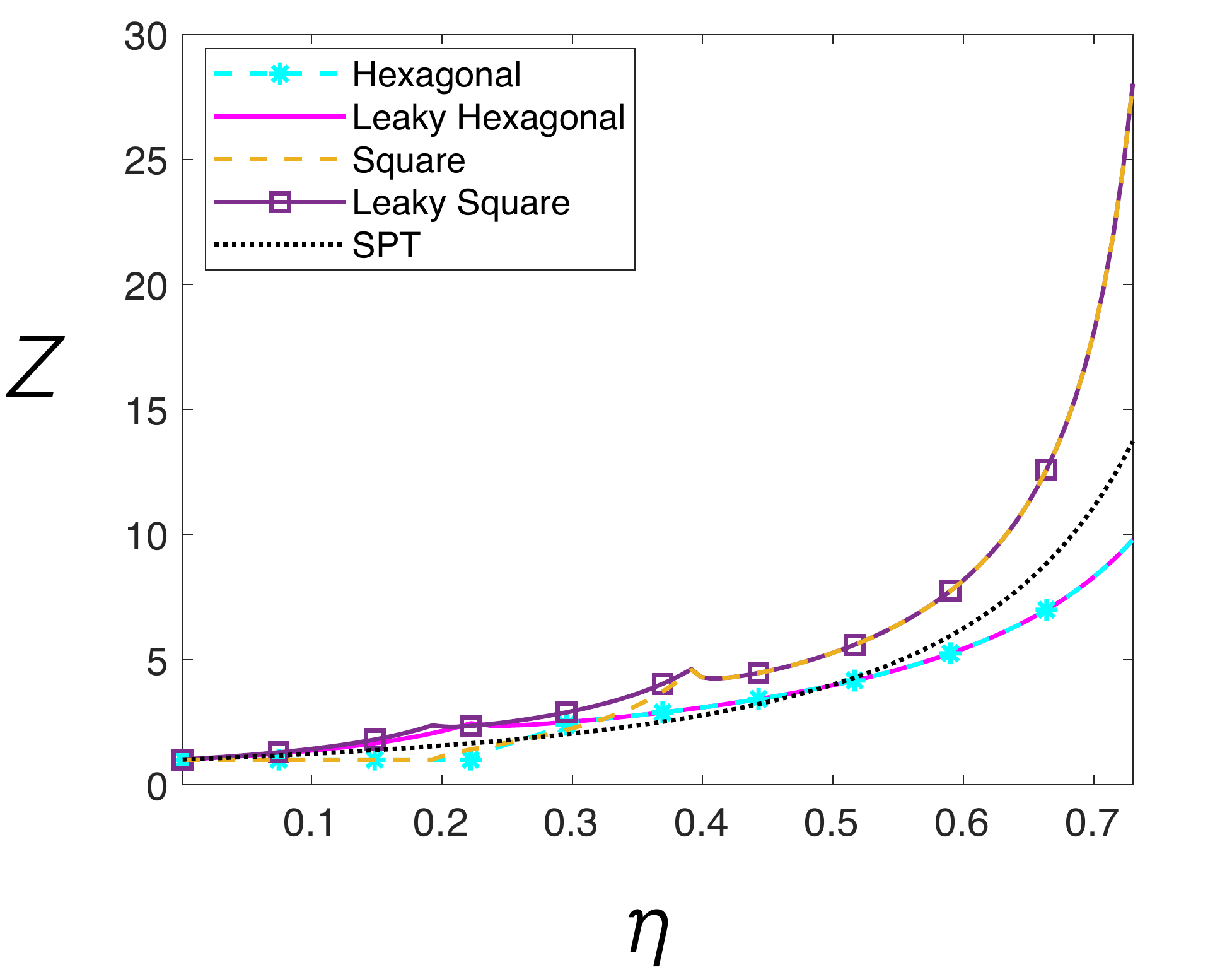}
         \caption{Comparison of 2D \edit{compressibility factors} on square and hexagonal lattices and the \edit{scaled particle theory (SPT)} equation of state.}
         \label{fig:comp2D}
\end{figure}

Figure \ref{fig:comp2D} shows the resulting \edit{compressibility factors} computed through Eq.~\eqref{eq:cf}. It is interesting to note that whereas the \edit{compressibility factor} for the hexagonal lattice is a smooth function of $\eta$, there is a kink in the square lattice \edit{compressibility factor} that arises from a discontinuity in the second derivative of $\mathcal{F}(v)$ at the leaky cell transition as well as a discontinuity at the percolation transition.

\subsection{Comparison to 2D \edit{scaled particle theory}}
Finally, we compare the \edit{compressibility factors} computed from the above formulas to the 2D \edit{scaled particle theory} equation of state:
\begin{equation}
Z(\eta) = \frac{1}{(1-\eta)^2}.
\end{equation}
As shown in Fig.~\ref{fig:comp2D}, we find that the hexagonal lattice \edit{compressibility factor} is in reasonable agreement with \edit{the scaled particle theory}, whereas the square lattice \edit{compressibility factor} is significantly different and has distinctive qualitative features, such as non-monotoniticity and a loss of smoothness at $\eta \approx 0.4$ and $\eta \approx 0.2$.

\section{\label{app:1D} Exact leaky model on 1D lattices: Tonks' gas}
Recall that for $N$ rods, each of length $\sigma$, confined within a segment of length $\ell$, we call
\begin{equation}\label{eq:lambda}
\lambda:=\frac{\ell}{N}
\end{equation}
the \emph{specific length} and
\begin{equation}\label{eq:theta}
\eta:=\frac{\sigma N}{\ell}=\frac{\sigma}{\lambda}
\end{equation}
is the packing fraction as before.
We imagine the segment partitioned in $N$ equal permeable cells, each of length $\lambda$. We denote by $x_j$ the coordinate of the center of mass of the $j$-th rod and, for a given parameter $0\leqq\alpha\leqq1$, we assume that 
\begin{align}
	\label{eq:x_j_bounds}
	(j-1)\lambda-\left(\alpha-\frac12\right)\sigma\leqq x_j&\leqq j\lambda+\left(\alpha-\frac12\right)\sigma, \notag \\ &\quad 2\leqq j\leqq N-1.
\end{align}
We call $\alpha$ the \emph{leak} parameter, as $\alpha\sigma$ is the maximum length each end-point of every rod can go outside the ideal restraining cell.

To compute the entropy $S_N$ of such a system, we start by computing the partition function $Z_N$, which is defined by the following integral,
\begin{align}
	\label{eq:partition_function}
	Z_N:=&\int_{(N-1)\lambda-\beta\sigma}^{\lambda N-\frac{\sigma}{2}}\int_{\lambda-\beta\sigma}^{\min\{x_3-\sigma,2\lambda+\beta\sigma\}} \notag\\
	&\dots\int_{\frac{\sigma}{2}}^{\min\{x_2-\sigma,\lambda+\beta\sigma\}}\ud x_1\ud x_2\dots\ud x_N,
\end{align}
where
\begin{equation}
	\label{eq:beta}
	\beta:=\alpha-\frac12.
\end{equation}
For a given configuration where $x_2,\dots,x_N$ are frozen, we compute the first of the nested integrals in \eqref{eq:partition_function} as
\begin{align}
	\label{eq:f_2}
	f_2(x_2)&:=\int_{\frac{\sigma}{2}}^{\min\{x_2-\sigma,\lambda+\beta\sigma\}}\ud x \notag\\
	&=
	\begin{cases}
		x_2-\frac32\sigma &\lambda-\beta\sigma\leqq x_2\leqq\lambda+(\beta+1)\sigma,\\
		\lambda+\left(\beta-\frac12\right)\sigma & \lambda+(\beta+1)\sigma\leqq x_2\leqq 2\lambda+\beta.
	\end{cases}
\end{align}
It turns out that, provided that
\begin{equation}
	\label{eq:validity_bound}
	\lambda\geqq2(\beta+1)\sigma,
\end{equation}
also $f_n$, the $(n-1)$-th of the nested integrals in \eqref{eq:partition_function}, is a piece-wise linear function of $x_n$,
\begin{align}
	\label{eq:f_n}
	&f_n(x)= \notag\\
	&\begin{cases}
		A_nx+B_n&(n-1)\lambda-\beta\sigma\leqq x\leqq(n-1)\lambda+(\beta+1)\sigma,\\
		C_n &(n-1)\lambda+(\beta+1)\sigma\leqq x\leqq n\lambda+\beta\sigma,
	\end{cases}
\end{align}
with
\begin{equation}
	\label{eq:C_n}
	C_n=A_n[(n-1)\lambda+(\beta+1)\sigma]+B_n.
\end{equation}
Moreover, the coefficients $A_n$ and $B_n$ satisfy the recurrence relation
\begin{equation}
	\label{eq:A_B_recursion}
	\begin{split}
	A_{n+1}&=C_n\\
	B_{n+1}&=(2\beta+1)\sigma B_n+\frac12(2\beta+1)\sigma[\sigma+2\lambda(n-1)]A_n \\
	&\quad-[(n-1)\lambda+(\beta+2)\sigma]C_n.
	\end{split}
\end{equation}
It follows from \eqref{eq:partition_function} that 
\begin{equation}
	\label{eq:Z_N}
	\begin{split}
	Z_N&=\int_{(N-1)\lambda-\beta\sigma}^{N\lambda-\frac{\sigma}{2}}f_N(x)\ud x\\&=\frac12(2\beta+1)\sigma[\sigma+2\lambda(N-1)]A_N+(2\beta+1)\sigma B_N \\
	&\quad+\left[\lambda-\left(\beta+\frac32\right)\sigma\right]C_N.
	\end{split}
\end{equation}
Combining \eqref{eq:Z_N}, \eqref{eq:A_B_recursion}, and \eqref{eq:C_n}, we arrive at
\begin{equation}
	\label{eq:C_Z_recursion}
	\begin{split}
	C_{N+1}&=Z_N+\left(\beta+\frac12\right)\sigma C_N,\\
	Z_{N+1}&=\left[\lambda+\left(\beta-\frac12 \right)\sigma\right]Z_N \\
	&\quad+\sigma\left[ \left(\beta+\frac12\right)\lambda-\left(2\beta+\beta^2+\frac34\right)\sigma\right]C_N,
	\end{split}
\end{equation}
which in matrix form reads as
\begin{equation}
	\begin{pmatrix}
		C_{N+1}\\Z_{N+1}
	\end{pmatrix}=
	\label{eq:recursion_matrix_form}
	\mathsf{A}
\begin{pmatrix}
	C_N\\Z_N
\end{pmatrix},
\end{equation}
where $\mathsf{A}$ is the matrix
\begin{equation}
	\label{eq:A_matrix}
	\mathsf{A}:=\begin{bmatrix}
		\left(\beta+\frac12\right)\sigma & 1\\
		\sigma\left[ \left(\beta+\frac12\right)\lambda-\left(2\beta+\beta^2+\frac34\right)\sigma\right]&\lambda+\left(\beta-\frac12\right)\sigma
	\end{bmatrix},
\end{equation}
which possesses two real positive eigenvalues,
\begin{align}\label{eq:A_eigenevalues}
	\mu_{1,2}=\frac12\left[\lambda+(2\alpha-1)\sigma\pm\sqrt{[\lambda+(2\alpha-1)\sigma]^2-8\alpha^2\sigma^2} \right],
\end{align}
where use has been made of \eqref{eq:beta}.
Letting
\begin{equation}
	\label{eq:A_eigenvectors}
	\begin{pmatrix}
		v_{11}\\v_{12}
	\end{pmatrix}
\quad\text{and}\quad
\begin{pmatrix}
	v_{21}\\v_{22}
\end{pmatrix}
\end{equation}
denote the corresponding eigenvectors, the solution of the recurrence relation \eqref{eq:recursion_matrix_form} with initial value
\begin{equation}
	\label{eq:recursion_initial_value}
	\begin{pmatrix}
		C_0\\Z_0
	\end{pmatrix}=
\frac12\begin{pmatrix}
	\frac{1}{\sigma}\\1
\end{pmatrix}
\end{equation}
is given by
\begin{equation}
	\label{eq:recursion_solution}
	\begin{split}
		C_N&=c_1 v_{11}\mu_1^N+c_2v_{21}\mu_2^N,\\
		Z_N&=c_1 v_{12}\mu_1^N+c_2v_{22}\mu_2^N,
	\end{split}
\end{equation}
where $c_1$ and $c_2$ are solutions of the linear system
\begin{equation}
	\label{eq:c_1_c_2}
	\begin{split}
		c_1 v_{11}+c_2v_{21}=\frac{1}{2\sigma},\\
		c_1 v_{12}+c_2v_{22}=\frac12,
	\end{split}
\end{equation}
with the normalization $v_{11}=v_{21}=1$.

Since 
\begin{equation}
	\label{eq:etntropy_formula}
	S_N=k_BT\ln Z_N,
\end{equation}
in the limit as $N\gg1$, $S_N$ is uniquely determined by the larger eigenvalue $\mu_2$ of $\mathsf{A}$,
\hspace*{-.2in}\vbox{\begin{align}
	\label{eq:entropy_thermodynamic_limit}
	&S_N=k_BTN\ln\mu_2 \notag \\
	&=k_BTN\ln\frac12\left[(\lambda+(2\alpha-1)\sigma)+\sqrt{[\lambda+(2\alpha-1)\sigma]^2-8\alpha^2\sigma^2} \right],
\end{align}}
where use has also been made of \eqref{eq:theta}. By \eqref{eq:validity_bound}, this formula is only valid for
\begin{equation}
	\label{eq:range_of_validity}
	0\leqq\eta\leqq\frac{1}{2\alpha+1}.
\end{equation}
For larger values of $\eta$, \eqref{eq:f_n} is no longer valid, as $f_n$ is delivered by a piece-wise quadratic function, which first becomes cubic, and then of ever increasing order as the upper bound of $\eta$ approaches $1$. In such cases the above computations become increasingly cumbersome. 

\bibliography{biblio_v2}

\end{document}